\documentclass[12pt]{article}
 \pdfoutput=1
 \usepackage[utf8]{inputenc}
\usepackage{amssymb}
\usepackage{amsmath}
\usepackage{amstext}
\usepackage{array}
\usepackage{graphicx,epsfig}
\usepackage{epsfig}
\usepackage{verbatim} 
\usepackage{fancybox}
\usepackage{color}
\usepackage{ulem}
\usepackage{enumitem}
\usepackage{subfigure}
\usepackage{bbm}
\usepackage{parskip}
\usepackage{cite}
\usepackage{tensor}
\usepackage{braket}
\usepackage{longtable}
\usepackage[margin=1cm]{caption}

\newcommand{\Comment}[1]{{}}
\definecolor{MyDarkBlue}{rgb}{0.15,0.15,0.45}
\usepackage[linktocpage=true]{hyperref}
\hypersetup{
colorlinks=true,
citecolor=MyDarkBlue,
linkcolor=MyDarkBlue,
urlcolor=MyDarkBlue,
pdfsubject={hep-th}
}

\definecolor{green3}{RGB}{44, 160, 44}

\newcommand{\be}{\begin{equation}}
\newcommand{\ee}{\end{equation}}
\newcommand{\bea}{\begin{eqnarray}}
\newcommand{\eea}{\end{eqnarray}}
\newcommand{\beas}{\begin{eqnarray*}}
\newcommand{\eeas}{\end{eqnarray*}}
\newcommand{\nn}{\nonumber}

\def\({\left(}
\def\){\right)}

\newcommand{\la}{\langle}
\newcommand{\ra}{\rangle}

\newcommand{\beq}{\begin{equation}}
\newcommand{\eeq}{\end{equation}}
\newcommand{\bal}{\begin{aligned}}
\newcommand{\eal}{\end{aligned}}

\newlength{\fdagwidth}
\newlength{\diagupwidth}
\newlength{\stepback}

\numberwithin{equation}{section}

\begin{document}


\begin{center}
{\Large \bf{${\cal N}=2$ Supersymmetric Partially Massless Fields  }}\\ \vspace{.2cm}
{\Large \bf{  and Non-Unitary Superconformal Representations}}\\ 
\end{center} 

\vspace{2truecm}

\thispagestyle{empty}
\centerline{{\large Noah Bittermann,${}^{\rm a,}$\footnote{\href{mailto:nb2778@columbia.edu}{\texttt{nb2778@columbia.edu}}}\,Sebastian Garcia-Saenz,${}^{\rm b,}$\footnote{\href{mailto:s.garcia-saenz@imperial.ac.uk}{\texttt{s.garcia-saenz@imperial.ac.uk}}} Kurt Hinterbichler,${}^{\rm c,}$\footnote{\href{mailto:kurt.hinterbichler@case.edu}{\texttt{kurt.hinterbichler@case.edu}}} 
Rachel A. Rosen${}^{\rm a,}$\footnote{\href{mailto:rar2172@columbia.edu}{\texttt{rar2172@columbia.edu}}}}}
\vspace{.7cm}

 \centerline{{\it ${}^{\rm a}$Center for Theoretical Physics, Department of Physics, }}
 \centerline{{\it Columbia University, New York, NY 10027}} 
 \vspace{.35cm}
 
 \centerline{{\it ${}^{\rm b}$Theoretical Physics, Blackett Laboratory, Imperial College, London, SW7 2AZ, U.K.}}
 \vspace{.35cm}

 \centerline{{\it ${}^{\rm c}$CERCA, Department of Physics, Case Western Reserve University, }}
 \centerline{{\it 10900 Euclid Ave, Cleveland, OH 44106}} 
 \vspace{.35cm}

\setcounter{footnote}{0}

\begin{abstract}
\vspace{0.0truecm}

We find and classify the simplest ${\cal N}=2$ SUSY multiplets on AdS$_4$ which contain partially massless fields.  We do this by studying representations of the ${\cal N}=2$, $d=3$ superconformal algebra of the boundary, including new shortening conditions that arise in the non-unitary regime.  Unlike the ${\cal N}=1$ case, the simplest ${\cal N}=2$ multiplet containing a partially massless spin-2 is short, containing several exotic fields.  More generally, we argue that ${\cal N}=2$ supersymmetry allows for short multiplets that contain partially massless spin-$s$ particles of depth $t=s-2$.

\end{abstract}

\newpage

\setcounter{tocdepth}{2}
\tableofcontents
\newpage

\section{Introduction}

Partially massless (PM) particles are exotic representations of the (anti) de Sitter ((A)dS) isometry algebras \cite{Deser:1983tm,Deser:1983mm,Higuchi:1986py,Brink:2000ag,Deser:2001pe,Deser:2001us,Deser:2001wx,Deser:2001xr,Zinoviev:2001dt,Skvortsov:2006at,Skvortsov:2009zu}.  These representations occur at discrete mass values relative to the (A)dS curvature and possess a gauge symmetry, despite their mass terms.  Partially massless fields exist for all spins $s \geq 1$ and are labeled by their depth $t$: for bosons $t \in\{0, 1, ...s - 1\}$ and for fermions $t \in\{1/2, 3/2, ...,s - 1\}$.  For a depth-$t$ PM field, the gauge symmetry removes the helicity components with helicities $\leq t$ from the particle.  {The depth $t=s-1$ field corresponds to the usual massless representation.} $\mathcal{N} = 1$ supersymmetric (SUSY) extensions of partially massless representations of AdS$_{4}$ were studied in \cite{Garcia-Saenz:2018wnw,Buchbinder:2019olk}.  In this work, we follow up those results by finding ${\cal N}=2$ SUSY multiplets on AdS$_4$ which contain partially massless fields.  We will also comment briefly on some general properties of PM representations expected for $\mathcal{N}>2$.

There are several reasons to extend the previous studies to ${\cal N}>1$. Given the importance of the AdS group and its supersymmetric extensions -- holography and higher-spin theory being two examples \cite{Bekaert:2013zya,Basile:2014wua,Alkalaev:2014nsa,Joung:2015jza,Brust:2016zns} -- a complete classification of the SUSY AdS representations is desirable.  Moreover, a deeper understanding of partially massless representations could also shed light on the difficulties encountered when constructing interacting field theories for PM particles ~\cite{Zinoviev:2006im,Hassan:2012gz,Hassan:2012rq,deRham:2012kf,Hassan:2013pca,Deser:2013uy,deRham:2013wv,Zinoviev:2014zka,Garcia-Saenz:2014cwa,Hinterbichler:2014xga,Joung:2014aba,Alexandrov:2014oda,Hassan:2015tba,Hinterbichler:2015nua,Cherney:2015jxp,Gwak:2015vfb,Gwak:2015jdo,Garcia-Saenz:2015mqi,Hinterbichler:2016fgl,Bonifacio:2016blz,Apolo:2016ort,Apolo:2016vkn,Bernard:2017tcg,Boulanger:2018dau,Joung:2019wwf}.  Going beyond the restrictions of unitarity is also desirable.  From the boundary conformal field theory (CFT) point of view, PM fields in AdS are dual to CFT currents satisfying higher derivative conservation conditions \cite{Dolan:2001ih}, which occur only in non-unitary CFTs.  Non-unitary CFTs and the representations they realize have found applications in condensed matter systems \cite{Maassarani:1996jn}, and in understanding the analytic structure of the conformal blocks in ordinary unitary CFTs \cite{Penedones:2015aga,Erramilli:2019njx,Sen:2018del}.    Unitary superconformal representations have been extensively studied and classified \cite{Flato:1983te,Dobrev:1985qv,Dobrev:1985vh,Dobrev:1985qz,Minwalla:1997ka,Dolan:2008vc,Bhattacharya:2008zy,Cordova:2016emh}, however the non-unitary cases remain relatively unexplored, though some results are known \cite{Oshima:2016gqy,Yamazaki:2016vqi,Sen:2018del,Garcia-Saenz:2018wnw,Yamazaki:2019yfd}.  Here we will uncover new phenomena for ${\cal N}=2$, such as exotic shortenings and extended modules, that occur in non-unitary regions.

Another reason to consider ${\cal N}=2$ is that when considering dS space, the supersymmetric extensions of the dS group require an even number of supercharges \cite{Pilch:1984aw,Lukierski:1984it}. The essential reason is that the supercharges of SUSY-dS$_4$ must satisfy a symplectic Majorana condition, and for this $\mathcal{N}$ must be even. {The equivalent statement at the Lagrangian level is that the generator of SUSY transformations cannot be a standard Majorana Killing spinor, since such spinors do not exist in dS$_4$ (see \cite{Anous:2014lia} for a related discussion).}  Thus, if one is interested in constructing supersymmetric theories with partially massless particles on dS (because the bosonic PM fields are themselves unitary on dS, but not on AdS), one needs to consider even ${\cal N}$.

As we will see, extended SUSY allows for a rich structure of multiplets containing PM fields. While for $\mathcal{N}=1$, PM fields always sit in long supermultiplets \cite{Garcia-Saenz:2018wnw}, there is no reason to expect that this will be the case for $\mathcal{N}\geq2$.  {By ``long'' here, we mean having as many conformal primary states as a generic long supermultiplet; the supermultiplet itself can contain conformal primaries which are themselves short states due to the presence of the PM  gauge symmetries, but this is unrelated to SUSY.}   Indeed, for $\mathcal{N}=2$ we find that supermultiplets with PM states can be either long or short, in addition to the exotic possibility of featuring a so-called ``extended module'' phenomenon.  In particular, we find that for $\mathcal{N}=2$ there are short supermultiplets where the highest-spin state in the multiplet is a spin-$s$ partially massless state of depth $t=s-2$.  Thus, unlike the ${\cal N}=1$ case, the simplest ${\cal N}=2$ supermultiplet containing a partially massless spin-2 particle is short, and it also contains several exotic fields.

\bigskip

\noindent {\bf Conventions}:  This paper relies heavily on the notation and concepts introduced in \cite{Garcia-Saenz:2018wnw}, and the conventions used here are detailed there.

\bigskip 
\section{The superconformal algebra\label{susycalgesecef}}

As in \cite{Garcia-Saenz:2018wnw}, we study supersymmetric extensions of the PM representations via the AdS/CFT correspondence.  We are interested in AdS$_{4}$ SUSY, thus we study $d = 3$ superconformal symmetry on the boundary.

\subsection{${\cal N}$ extended $d=3$ superconformal algebra}

The generators of the euclidean ${\cal N}$ extended superconformal algebra are 
\bea &&  P^i,\  J^{ij},\   D,\ \ K^i,  \    Q^{aI},\  S_{a }^I,\ R^{IJ} \,.
\eea 
The $P^i$ are the translations and the anti-symmetric $J^{ij}$ are the rotations, which together generate the Poincare transformations of $d=3$ Euclidean space.  The dilation is $D$ and the special conformal generators are $K^i$, which together with the Poincare generators generate the conformal symmetries.  $Q^{aI}$ are the spinor-valued supersymmetries, labelled by the index $I=1,\ldots, {\cal N}$.   Together with the Poincare generators they generate ${\cal N}$ extended SUSY.  The $S_{a}^{I}$ are the special superconformal generators, and the anti-symmetric $R^{IJ}$ are $so({\cal N})$ $R$-symmetries, which together complete the SUSY generators and conformal generators into to the ${\cal N}$ extended superconformal algebra.

The non-vanishing (anti)commutators are as follows \cite{Minwalla:1997ka}: 
First there are the usual commutators of the Poincare algebra,
\bea \left[J^{ij},P^k\right] &=& i\left(-\delta^{ki} P^j+\delta^{kj} P^i\right)\,  ,  \nn \\
\left[J^{ij},J^{kl}\right]&=&i\left(-\delta^{ik}J^{jl}+\delta^{jk}J^{il}-\delta^{jl}J^{ik}+\delta^{il}J^{jk} \right)\,. \label{pvectorc}
\eea 
The commutators which when taken together with \eqref{pvectorc} fill out the conformal algebra are
\bea \left[D,P^i\right]&=&  P^i \, ,  \label{weightp1} \nn\\ 
 \left[D,K^i\right]&=& - K^i \, ,\label{weightkm1}\nn\\
  \left[K^i,P^j\right]&=&2(\delta^{ij} D+iJ^{ij})\, ,\nn\\
   \left[J^{ij},K^k\right]&=& i\left(-\delta^{ki}K^j+\delta^{kj}K^i\right)\, .\label{kvectorc}
   \eea 
The commutators which when taken together with \eqref{pvectorc} form the ${\cal N}$ extended SUSY algebra are
\bea && \left\{Q^{aI},Q^{bJ}\right\}=2\sigma_i^{ab}P^i\,\delta^{IJ} , \label{mainsusycome} \\ 
&&  \left[ J_{ij},Q^{aI}\right]=-{i\over 2}\left(\sigma_{ij}\right)^a_{\ b}Q^{bI},\label{qisspinore}\\
&& \left[R^{IJ},R^{KL}\right]=i\left(-\delta^{IK}R^{JL}+\delta^{J K}R^{IL}-\delta^{JL}R^{IK}+\delta^{IL}R^{JK} \right)  , \label{mainrsymcomme}\\
&& \left[R^{IJ},Q^{a K}\right]=i\left(-\delta^{IK}Q^{aJ} +\delta^{JK}Q^{aI}   \right)  .\label{Qisvectorunderre}
 \label{mainsssyq}
\eea
The first line  \eqref{mainsusycome} is the main anti-commutator indicative of SUSY, \eqref{qisspinore} shows that $Q^{aI}$ transforms as a spinor under rotations, \eqref{mainrsymcomme} is the statement that $R^{IJ}$ forms an $so({\cal N})$, and \eqref{Qisvectorunderre} shows that $Q^{aI}$ transforms as a vector under this $so({\cal N})$.

The remaining non-trivial commutators, which when taken with the above fill out the superconformal algebra, are
\bea && \left[ J_{ij},S^{aI}\right]=-{i\over 2}\left(\sigma_{ij}\right)^a_{\ b}S^{bI} ,\ \  \left[R^{IJ},S^{a K}\right]=i\left(-\delta^{IK}S^{aJ} +\delta^{JK}S^{aI}   \right) ,\nn\\
&& \left[ D, Q^{a I}\right]={1\over 2}Q^{a I}\, ,\ \   \left[ D, S^{a I}\right]=-{1\over 2}S^{a I}\,\nn\\
&&  \left\{S^{aI},S^{bJ}\right\}=-2\sigma_i^{ab}K^i\,\delta^{IJ} , \label{mainsssys}\nn\\
&&   \left\{Q^{aI},S^{bJ}\right\}=2\delta^{IJ}\epsilon^{ab}D-i\delta^{IJ}\sigma_{ij}^{ab}J^{ij}+2i\epsilon^{ab}R^{IJ} ,\label{mainsccomae} \nn\\
&& \left[K_i,Q^{aI}\right]=-\left(\sigma_i\right)^a_{\ b}S^{bI} ,\nn\\
&&   \left[P_i,S^{aI}\right]=\left(\sigma_i\right)^a_{\ b}Q^{bI} .\label{remainingsuscyecae}
\eea
The first line of \eqref{remainingsuscyecae} indicates that $S^{a I}$ transforms as a spinor under rotations and a vector under  $so({\cal N})$ $R$-symmetry,  the second line indicates that $Q^{a I}$ carries scaling dimension $1/2$ and $S_{a }^I$ carries scaling dimension $-1/2$.

In radial quantization, the generators satisfy the conjugation relations\footnote{Recall that with our conventions, outlined in \cite{Garcia-Saenz:2018wnw}, there is a subtlety with the indices; the condition ${Q^{a}}^\dag=S_{a}$ implies ${S_{a}}^\dag=Q^{a}$, but when both raising and lowering indices we get a sign: ${Q_{a}}^\dag=-S^{a},\   {S^{a}}^\dag=-Q_{a}\,.$}
\be P^{i \dag}=K^i,\ \ D^\dag=D,\ \ \ \ J^{ij \dag}=J^{ij},\ \ \ {Q^{aI}}^\dag=S_{a}^I,\ \ \ R^{IJ\dag}=R^{IJ}  \,. \label{PDJcong}\ee  The (anti)commutation relations above are all consistent with the reality conditions \eqref{PDJcong}.

\subsection{Algebra in spinor form}

It will be convenient to put the algebra into pure spinor form by contracting the various 3$d$ vectors with sigma matrices.  The translations and special conformal generators become symmetric 2-index spinors,
\be P^{ab}=\sigma^{ab}_i P^i,\ \ \ \ P^i=-{1\over 2}\sigma_{ab}^i P^{ab},\ee
\be K^{ab}=\sigma^{ab}_i K^i,\ \ \ \ K^i=-{1\over 2}\sigma_{ab}^i K^{ab},\ee
and the rotations are dualized into a vector and then converted into a symmetric 2-index spinor,
\be  J^{i}=-{1\over 2}\epsilon^{ijk}J_{jk},\ \ \ J^{ij}=-\epsilon^{ijk}J_k,\ \ \ J^{ab}=\sigma^{ab}_i J^i,\ \ \ \ J^i=-{1\over 2}\sigma_{ab}^i J^{ab}.\ee

The commutators now take the form
\bea 
\left[J^{ab},P^{cd}\right]&=&{1\over2}\left(\epsilon^{ac}P^{bd}+\epsilon^{bc}P^{ad}+\epsilon^{ad}P^{cb}+\epsilon^{bd}P^{ca}\right),\nn\\
 \left[J^{ab},J^{cd}\right]&=&{1\over2}\left(\epsilon^{ac}J^{bd}+\epsilon^{bc}J^{ad}+\epsilon^{ad}J^{cb}+\epsilon^{bd}J^{ca}\right), 
 \eea
 \bea
 \left[D,P^{ab}\right]&=&  P^{ab},\ \ \  \left[D,K^{ab}\right]= - K^{ab}\,, \nn\\
    \left[K^{ab},P^{cd}\right]&=&-2\left(\epsilon^{ac}\epsilon^{bd}+\epsilon^{bc}\epsilon^{ad}\right)D-\left(\epsilon^{ac}J^{bd}+\epsilon^{bc}J^{ad}+\epsilon^{ad}J^{cb}+\epsilon^{bd}J^{ca}\right)\, ,\nn \\
  \left[J^{ab},K^{cd}\right]&=&{1\over2}\left(\epsilon^{ac}K^{bd}+\epsilon^{bc}K^{ad}+\epsilon^{ad}K^{cb}+\epsilon^{bd}K^{ca}\right),
   \eea 
\bea &&\left[J^{ab},Q^{cI}\right]={1\over 2}\left(\epsilon^{ac}Q^{bI}+\epsilon^{bc}Q^{aI}\right),\ \ \ \left[R^{IJ},Q^{a K}\right]=i\left(-\delta^{IK}Q^{aJ} +\delta^{JK}Q^{aI}   \right)\, ,\nn\\
&&  \left[R^{IJ},R^{KL}\right]=i\left(-\delta^{IK}R^{JL}+\delta^{J K}R^{IL}-\delta^{JL}R^{IK}+\delta^{IL}R^{JK} \right), \nn \\
&&  \left\{Q^{aI},Q^{bJ}\right\}=2P^{ab}\,\delta^{IJ} \, ,
\eea
\bea && \left[J^{ab},S^{cI}\right]={1\over 2}\left(\epsilon^{ac}S^{bI}+\epsilon^{bc}S^{aI}\right),\ \ \  \left[R^{IJ},S^{a K}\right]=i\left(-\delta^{IK}S^{aJ} +\delta^{JK}S^{aI}   \right)\, ,\nn \\  
&& \left[ D, Q^{aI}\right]={1\over 2}Q^{aI},\ \ \ \left[ D, S^{aI}\right]=-{1\over 2}S^{aI}\, ,\nn \\
&& \left\{S^{aI},S^{bJ}\right\}=-2K^{ab}\,\delta^{IJ}, \nn\\
&& \left\{Q^{aI},S^{bJ}\right\}=2\delta^{IJ}\epsilon^{ab}D-2\delta^{IJ}J^{ab}+2i\epsilon^{ab}R^{IJ},\nn\\
&&  \left[K^{ab},Q^{cI}\right]=\epsilon^{ac}S^{bI}+\epsilon^{bc}S^{aI},\ \ \ \left[P^{ab},S^{cI}\right]=-\left(\epsilon^{ac}Q^{bI}+\epsilon^{bc}Q^{aI}\right)\,.
\eea

The conjugation relations \eqref{PDJcong} now read
\bea && P^{ab\dag}=-K_{ab},\ \ \ D^\dag=D,\ \ \ J^{ab\dag}=-J_{ab},\ \ \ {Q^{aI}}^\dag=S_{a}^I,\ \ \ R^{IJ\dag}=R^{IJ}.
\eea

\subsection{${\cal N}=2$}

For ${\cal N}=2$, the case of primary interest for us, we have only one $R$ symmetry generator, $R\equiv R^{12}$, generating $u(1)_R$, and two sets of supercharges $Q^{a1}$, $Q^{a2}$, $S^{1}_a$, $S_a^{2}$, which transform under the $U(1)_R$ as 
\be \left[R,Q^{a 1}\right]=-iQ^{a2} \,,\ \  \left[R,Q^{a 2}\right]=iQ^{a1}\,,\ \  \left[R,S^{a 1}\right]=-iS^{a2} \,,\ \  \left[R,S^{a 2}\right]=iS^{a1}\, . \label{u1rfirstactione}\ee
We can diagonalize this $u(1)_R$ action by defining the linear combinations
\bea 
&& Q^a={1\over \sqrt{2}}\left(Q^{a 1}-i Q^{a 2}\right),\ \ \bar Q^a={1\over \sqrt{2}}\left(Q^{a 1}+i Q^{a 2}\right),\nn \\  
&& S^a={1\over \sqrt{2}}\left(S^{a 1}-i S^{a 2}\right),\ \ \bar S^a={1\over \sqrt{2}}\left(S^{a 1}+i S^{a 2}\right),  
\eea
in terms of which \eqref{u1rfirstactione} becomes
\be \left[R,Q^{a}\right]=Q^{a},\ \ \ \ \left[R,\bar Q^{a}\right]=-\bar Q^{a}\, ,\ \ \ \  \left[R,S^{a}\right]=S^{a},\ \ \ \ \left[R,\bar S^{a}\right]=-\bar S^{a}\, .\ee
The conjugation rules are now
\be Q^{a\dag}=\bar S_a,\ \ \ \bar Q^{a\dag}=S_a,\ \ \ S^{a\dag}=-\bar Q_a,\ \ \ \bar S^{a\dag}=-Q_a.\ee
The non-vanishing commutators involving more than one $Q$ and/or $S$ are now
\be \left\{Q^{a},\bar Q^{b}\right\}=2P^{ab}\,,\ \   \left\{S^{a},\bar S^{b}\right\}=-2K^{ab}\, ,\ \label{mainsssyqs2n2e}\ee
\be   \left\{Q^{a},\bar S^{b}\right\}=2\epsilon^{ab}(D-R)-2J^{ab},\ \ \  \left\{\bar Q^{a}, S^{b}\right\}=2\epsilon^{ab}(D+R)-2J^{ab}\, .\label{mainsccomaes2n22e} \ee

\bigskip
\section{${\cal N}=2$ Superconformal representations\label{susconforepss}}

We will be interested in finding representations of ${\cal N}=2$ SUSY which contain partially massless fields.  The various bosonic and fermionic PM fields on AdS$_{4}$ and their dual operators in CFT$_3$ are reviewed in Section 2 of \cite{Garcia-Saenz:2018wnw}, whose notation we follow.   Let us recall here the AdS/CFT mass formula and PM mass values.  A spin-$s$ fields in AdS$_{4}$ with mass $m$ correspond to spin-$s$ primary operators on the boundary with scaling dimension $\Delta$ related by
\begin{equation}
    m^2 L^2 = 
    \begin{cases}
        \Delta(\Delta - 2) & s = 0\\
        (\Delta + s - 2)(\Delta - s - 1) & s\geq \frac{1}{2}
    \end{cases}
    \label{mass2dim}
\end{equation}
where $L$ is the AdS radius.  There are two choices of boundary conditions for the AdS fields, for the ``standard quantization" boundary conditions one takes the greater root with $\Delta>\frac{3}{2}$, whereas for the ``alternate quantization" \cite{Klebanov:1999tb} boundary conditions one takes the lesser root with $\Delta<\frac{3}{2}$.  For bosons and fermions, the PM mass values are
\begin{equation}
    m^2_{s,t} = \frac{1}{L^2}(t - s + 1)(s + t)\hspace{1 cm}
\end{equation}
In the standard quantization, this means that both PM bosons and fermions of spin $s$ and depth $t$ have a scaling dimension of  $\Delta_{s, t} = t + 2.$

Superconformal representations can be constructed by joining together conformal representations. Conformal representations are built from a conformal primary.  A spin-$s$ conformal primary of weight $\Delta$ is indicated by $\ket{\Delta}^{a_1...a_{2s}}$. It has  $2s$ fully symmetric spinor indices and transforms under rotations as an irreducible representation of spin $s$,
\begin{align}
    &J_{ij}\ket{\Delta}^{a_1...a_{2s}} = -\frac{i}{2}(\sigma_{ij})\indices{^{a_1} _{b}}\ket{\Delta, r}^{ba_2...a_{2s}} - ... - \frac{i}{2}(\sigma_{ij})\indices{^{a_{2s}} _{b}}\ket{\Delta}^{a_1...a_{2s - 1}b} \,,\\
    &J_{ij}J^{ij}\ket{\Delta}^{a_1...a_{2s}} = 2s(s + 1)\ket{\Delta}^{a_1...a_{2s}}.
\end{align}
In terms of the rotation generators in spinor form this becomes
\be J^{ab}|\Delta\ra^{c_1\cdots c_{2s}}={1\over 2}\epsilon^{ac_1}|\Delta\ra^{bc_2\cdots c_{2s}}+{1\over 2}\epsilon^{ac_2}|\Delta\ra^{c_1bc_3\cdots c_{2s}}+\cdots+{1\over 2}\epsilon^{ac_{2s}}|\Delta\ra^{c_1\cdots c_{2s-1}b}+(a\leftrightarrow b)\ee
\be J^{ab}J_{ab}|\Delta\ra^{c_1\cdots c_{2s}}= -J^{ij}J_{ij}|\Delta\ra^{c_1\cdots c_{2s}}= -2J^{i}J_{i}|\Delta\ra^{c_1\cdots c_{2s}}= -2s(s+1)|\Delta\ra^{c_1\cdots c_{2s}}.\ee
Under dilations it has eigenvalue $\Delta$,
\be   D\ket{\Delta}^{a_1...a_{2s}}=\Delta  \ket{\Delta}^{a_1...a_{2s}}. \ee

A conformal primary obeys
\begin{equation}
   K_{i}\ket{\Delta}^{a_1...a_{2s}}=0
\end{equation}
and the rest of the representation is built by repeatedly acting with $P^{i}$.   A state with $l$ actions of $P^i$ has dimension $\Delta+l$, the states of this dimension are
\begin{equation}
    P^{i_1}\cdots P^{i_l}\ket{\Delta}^{a_1...a_{2s}}\,.
\end{equation}
These states are then further decomposed into irreducible representations under $J_{ij}$ by symmetrizing, removing traces, etc.  We will label a conformal multiplet whose primary has scaling dimension $\Delta$ and spin $s$ by $[s]_{\Delta}$.

Superconformal representations are constructed by joining together conformal representations via the action of the $Q$'s and the $S$'s.  We begin with a superconformal primary $\ket{\Delta, r}^{a_1...a_{2s}}$ which is a spin-$s$ conformal primary with conformal weight $\Delta$, which is also an eigenstate of the $R$-symmetry generator,
\begin{equation}
    R\ket{\Delta, r}^{a_1...a_{2s}} = r\ket{\Delta, r}^{a_1...a_{2s}}, \label{scprsymmee}
\end{equation}
and which is also annihilated by the $S$'s,
\begin{align}
    &S_{a}\ket{\Delta, r}^{a_1...a_{2s}} = 0, \,\,\,\bar{S}_{a}\ket{\Delta, r}^{a_1...a_{2s}} = 0.
\end{align}
We find the other conformal primaries within the superconformal multiplet by computing the $Q$ descendants of the superconformal primary, which come from acting with $Q$ (or $\bar{Q}$).  This raises $\Delta$ by $1/2$ and raises (lowers) $r$ by $1$.  This can be repeated until we get up to four total factors of $Q$, $\bar Q$.  After this the process terminates, because adding an additional $Q$ or $\bar Q$ reduces, after using the commutation relations, to acting with $P^{i}$ or causes the state to vanish.  In many instances, to actually construct a conformal primary from a given $Q$ descendent, we must also add on linear combinations of $P^{i}$ descendants so that the total state is annihilated by $K^{i}$. These $P$ descendants must have the same quantum numbers as the $Q$-descendants we are adding them to.  In this way, we construct a superconformal multiplet, which we label by $\{s\}_{\Delta, r}$.

In what follows we will list the conformal primaries present at each level for a given superconformal multiplet.  The $\{0\}_{\Delta, r}$ and $\{\frac{1}{2}\}_{\Delta, r}$ multiplets are qualitatively different from the multiplets for generic spin $s>1/2$, so we will present those first.  This is because acting with certain combinations of $Q$ lowers the spin; at most, the spin is lowered by $1$, and such states do not exist in the spin-$0$ and spin-$\frac{1}{2}$ multiplets.  The results for the $\{0\}_{\Delta, r}$ multiplet, the $\{\frac{1}{2}\}_{\Delta, r}$ multiplet, and the generic $\{s\}_{\Delta; r}$ multiplet are recorded in Table \ref{s0conformalprimary}, Table \ref{s1/2conformalprimary}, and Table \ref{sconformalprimary} respectively.   Note that there is an ambiguity in the ordering of the $Q$'s and $\bar Q$'s when acting on the superconformal primary, and different choices will lead to differed $P$ terms in the expressions.  Our choice is that in states with $\delta r\leq 0$, where $\delta r$ is the difference between the state's $r$ charge and the superconformal primary's $r$ charge, we have the $Q$'s acting first, and states with $\delta r>0$ are reversed from those with $\delta r<0$ by taking $Q\rightarrow\bar{Q}, \,\,\bar{Q}\rightarrow Q$ and $r\rightarrow-r$.   

For each conformal primary, there is a $CPT$-reversed state which can be constructed by carrying out $Q\rightarrow\bar{Q}, \,\,\bar{Q}\rightarrow Q$ and $r\rightarrow-r$.  When constructing Lagrangians or $CPT$ invariant theories, all multiplets must appear with their $CPT$ conjugates.   In all the multiplets that follow, the masses of the various bulk fields on the AdS side can be found from $\Delta$ through the AdS/CFT mass formula \eqref{mass2dim}.

\subsection{$s=0$}
\begin{figure}[h!]
\begin{center}
\epsfig{file=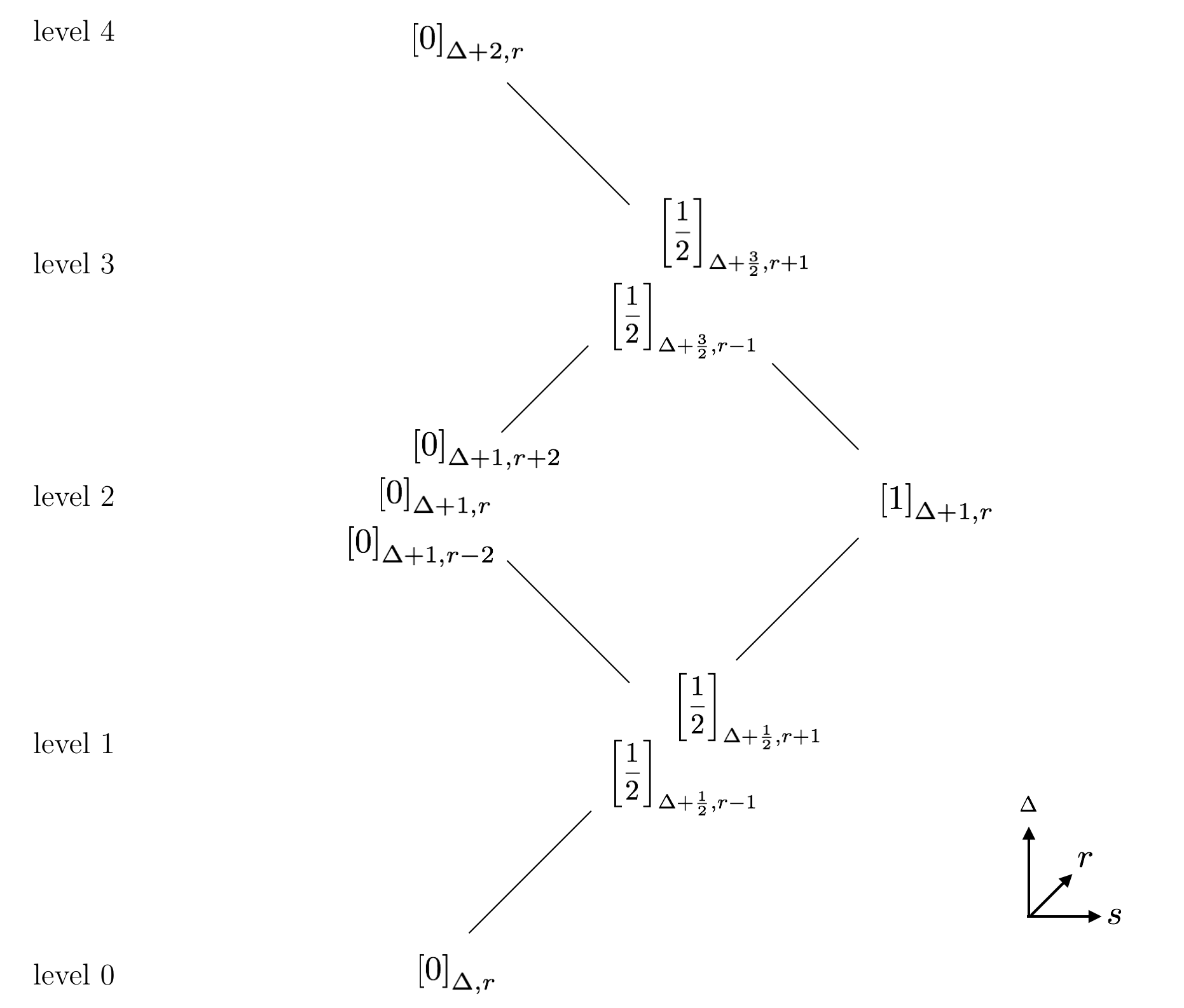,width=5.0in}
\caption{\small Generic scalar multiplet $\left\{ 0\right\}_{\Delta,r}$.  The notation $[s]_{\Delta, k}$ denotes each conformal primary within the superconformal multiplet.  The $[0]_{\Delta, k}$ conformal primary at level zero is also the superconformal primary.  There are four levels, which corresponds to the fact that we can act on the conformal primary with at most four supercharges $Q_{a}, \bar{Q}_{a}$.}
\label{s0multipletgeneric}
\end{center}
\end{figure}

The conformal primaries of the $\{0\}_{\Delta, r}$ multiplet are shown in Table \ref{s0conformalprimary}.  The multiplet can be visualized as in Figure \ref{s0multipletgeneric}.  From the bulk point of view, for generic values of $\Delta$ this multiplet contains 5 massive scalars (1 degree of freedom each) and a massive vector (3 degrees of freedom), for a total of 8 bosonic propagating degrees of freedom, and 4 massive spin $1/2$ fermions (2 degrees of freedom each), for a total of 8 fermionic propagating degrees of freedom.

\bigskip
\begin{center}
\begin{longtable}{ |m{3.25 cm}| m{11.5 cm}| }
\caption{Conformal primaries in the $\{0\}_{\Delta, r}$ superconformal multiplet.  
}
\label{s0conformalprimary}\\
\hline
\multicolumn{2}{|c|}{\textbf{Conformal Primaries of $\{0\}_{\Delta,r}$}}\\
\hline
\multicolumn{2}{|c|}{\textbf{Level 1}}\\
\hline 
$\left[ \frac{1}{2}\right]_{\Delta + \frac{1}{2},r- 1}$  & $\bar{Q}^{a_1}\ket{\Delta, r}$ \\
\hline
  $\left[ \frac{1}{2}\right]_{\Delta + \frac{1}{2},r+ 1}$  & $Q^{a_1}\ket{\Delta, r}$ \\
\hline
\multicolumn{2}{|c|}{\textbf{Level 2}}\\
 \hline
 $\left[0\right]_{\Delta +1,r }$ & $\bar{Q}^{c}Q_{c}\ket{\Delta, r}$ \\
\hline
 $\left[0\right]_{\Delta +1,r-2 }$ & $\bar{Q}^{c}\bar{Q}_{c}\ket{\Delta, r}$\\
\hline
 $\left[0\right]_{\Delta +1,r+2 }$ & $Q^{c}Q_{c}\ket{\Delta, r}$\\
\hline
 $\left[1\right]_{\Delta +1,r }$ & $\bar{Q}^{(a_1}Q^{a_2)}\ket{\Delta, r}  - \Big(\frac{\Delta - r}{\Delta}\Big)P^{a_1 a_2}\ket{\Delta, r}$\\
\hline
\multicolumn{2}{|c|}{\textbf{Level 3}}\\
\hline
 $\left[ \frac{1}{2}\right]_{\Delta + \frac{3}{2},r- 1}$ & $\bar{Q}^{c}\bar{Q}_{c}Q^{a_1}\ket{\Delta, r}-2\Big(\frac{\Delta - r}{\Delta - \frac{1}{2}}\Big) P\indices{^{a_1} _c}\bar{Q}^{c}\ket{\Delta, r}$\\
\hline
 $\left[ \frac{1}{2}\right]_{\Delta + \frac{3}{2},r+ 1}$ & $Q^{c}Q_{c}\bar{Q}^{a_1}\ket{\Delta, r}-2\Big(\frac{\Delta + r}{\Delta - \frac{1}{2}}\Big) P\indices{^{a_1} _c}Q^{c}\ket{\Delta, r}$\\
\hline
\multicolumn{2}{|c|}{\textbf{Level 4}}\\
\hline
 $\left[0\right]_{\Delta +2,r }$  & \vspace{-1cm} {\begin{align}\bar{Q}^{c}\bar{Q}_{c}Q^{d}Q_{d}\ket{\Delta, r}
 & +4\Big(\tfrac{\Delta -r-1}{\Delta -1}\Big)P_{cd}\bar{Q}^{(c}Q^{d)}\ket{\Delta, r}   \nonumber\\   & -2\tfrac{(\Delta - r) (\Delta -r-1)}{(\Delta -1) ( \Delta -{1\over 2})}P_{cd}P^{cd}\ket{\Delta, r}  \nonumber\end{align}} \vspace{-1cm} \\
\hline
\end{longtable}
\end{center}

\newpage

\clearpage

\subsection{$s=1/2$}
\begin{figure}[h!]
\begin{center}
\epsfig{file=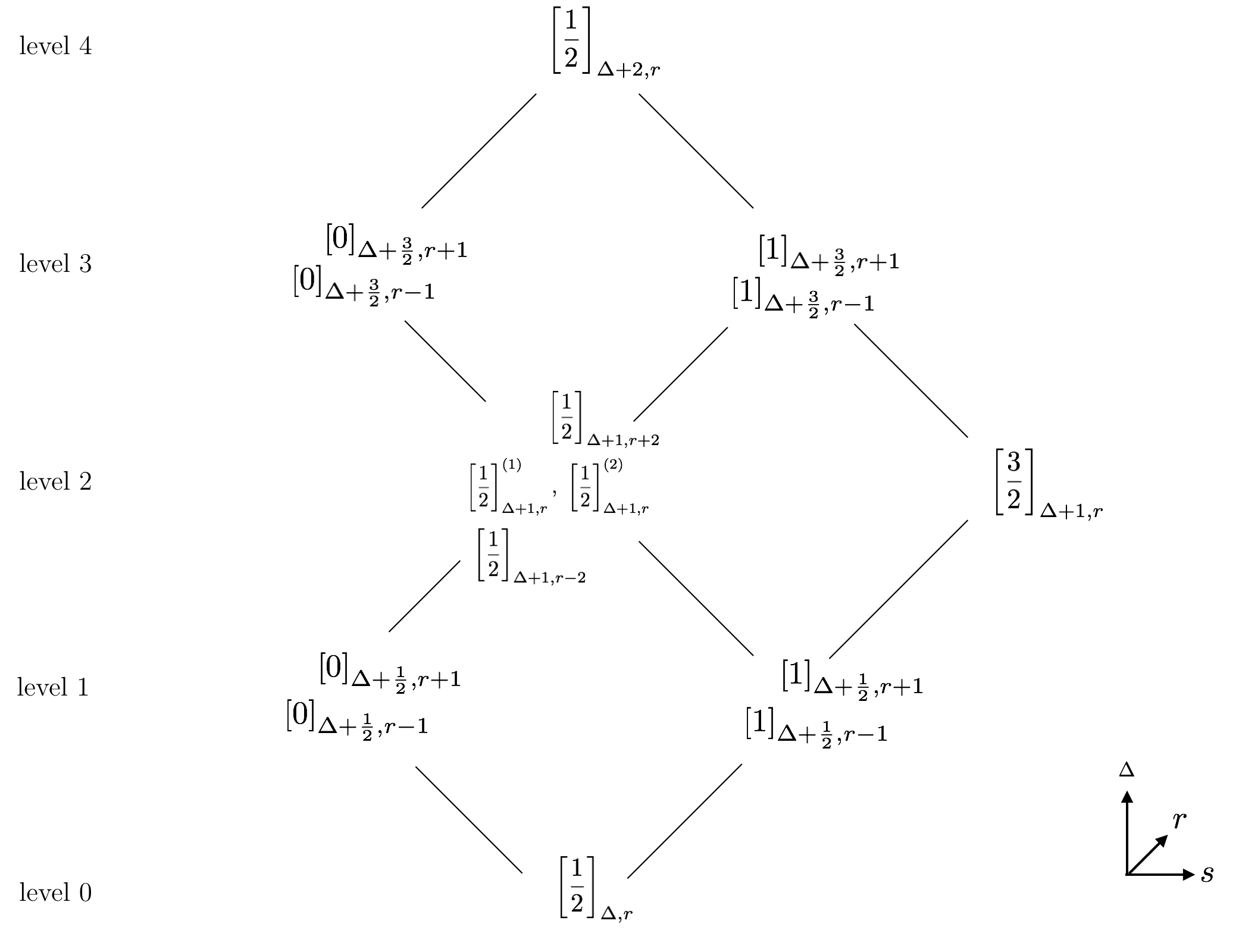,width=6.0in}
\caption{\small Conformal primaries of the generic multiplet $\left\{ {1\over 2}\right\}_{\Delta,r}$.  The $[{1\over 2}]_{\Delta, k}$ conformal primary at level zero is also the superconformal primary.}
\label{s1halfmultipletgeneric}
\end{center}
\end{figure}

The conformal primaries of the $\{\frac{1}{2}\}_{\Delta, r}$ multiplet are shown in Table \ref{s1/2conformalprimary}.  Note that at level 2 there are two different states with spin $1/2$, weight $\Delta+1$, and charge $r$, these are denoted by $[\frac{1}{2}]^{(1)}_{\Delta + 1, r}$ and $[\frac{1}{2}]^{(2)}_{\Delta + 1, r}$.  This multiplet can be visualized as in Figure \ref{s1halfmultipletgeneric}.  From the bulk point of view, for generic values of $\Delta$ this multiplet contains 4 massive scalars (1 degree of freedom each) and 4 massive vectors (3 degrees of freedom each), for a total of 16 bosonic propagating degrees of freedom, and 6 massive spin $1/2$ fermions (2 degrees of freedom each) and one massive spin $3/2$ fermion (4 degrees of freedom), for a total of 16 fermionic propagating degrees of freedom.

\bigskip
\begin{center}
\begin{longtable}{ |m{3.25 cm}| m{11.5 cm}| }
\caption{Conformal primaries in the $\{\frac{1}{2}\}_{\Delta, r}$ superconformal multiplet.}
\label{s1/2conformalprimary}\\
\hline
\multicolumn{2}{|c|}{\textbf{Conformal Primaries of  $\{\frac{1}{2}\}_{\Delta,r}$}}\\
\hline
\multicolumn{2}{|c|}{\textbf{Level 1}}\\
\hline
$[0]_{\Delta + \frac{1}{2}, r-1}$ & $\bar{Q}_{c}\ket{\Delta, r}^{c}$\\
\hline
$[0]_{\Delta + \frac{1}{2}, r+1}$ & $Q_{c}\ket{\Delta, r}^{c}$\\
\hline
$[1]_{\Delta + \frac{1}{2}, r - 1}$ & $\bar{Q}^{(a_1}\ket{\Delta, r}^{a_2)}$ \\
\hline
$[1]_{\Delta + \frac{1}{2}, r + 1}$ & $Q^{(a_1}\ket{\Delta, r}^{a_2)}$ \\
\hline
\multicolumn{2}{|c|}{\textbf{Level 2}}\\
\hline
$[\frac{1}{2}]^{(1)}_{\Delta + 1, r}$ & $\bar{Q}^{c}Q_{c}\ket{\Delta, r}^{a_1} + \Big(\frac{1}{\Delta - 1}\Big)P\indices{^{a_1}_c}\ket{\Delta, r}^{c}$ \\
\hline
$[\frac{1}{2}]^{(2)}_{\Delta + 1, r}$& $\bar{Q}_{c}Q^{a_1}\ket{\Delta, r}^{c} - \Big(\frac{\Delta - r - \frac{1}{2}}{\Delta - 1}\Big)P\indices{^{a_1}_c}\ket{\Delta, r}^{c}$\\
\hline
$[\frac{1}{2}]_{\Delta + 1, r - 2}$ & $\bar{Q}^{c}\bar{Q}_{c}\ket{\Delta, r}^{a_1}$\\
\hline
$[\frac{1}{2}]_{\Delta + 1, r + 2}$ & $Q^{c}Q_{c}\ket{\Delta, r}^{a_1}$\\
\hline
$[\frac{3}{2}]_{\Delta + 1, r}$ & $\bar{Q}^{(a_1}Q^{a_2}\ket{\Delta, r}^{a_3)}  - \Big(\frac{\Delta - r + \frac{1}{2}}{\Delta + \frac{1}{2}}\Big)P^{(a_1 a_2}\ket{\Delta, r}^{a_3)}$\\
\hline
\newpage
\hline
\multicolumn{2}{|c|}{\textbf{Level 3}}\\
\hline
$[0]_{\Delta + \frac{3}{2}, r - 1}$ &$\bar{Q}^{c}\bar{Q}_{c}Q_{d}\ket{\Delta, r}^{d}-2\Big(\frac{\Delta - r - \frac{3}{2}}{\Delta - \frac{3}{2}}\Big) P_{cd}\bar{Q}^{(c}\ket{\Delta, r}^{d)}$\\
\hline
$[0]_{\Delta + \frac{3}{2}, r + 1}$ &  $Q^{c}Q_{c}\bar{Q}_{d}\ket{\Delta, r}^{d}-2\Big(\frac{\Delta + r - \frac{3}{2}}{\Delta - \frac{3}{2}}\Big) P_{cd}Q^{(c}\ket{\Delta, r}^{d)}$\\
\hline
$[1]_{\Delta + \frac{3}{2}, r - 1}$& \vspace{-1cm}
{\begin{align}\bar{Q}^{c}\bar{Q}_{c}Q^{(a_1}\ket{\Delta, r}^{a_2)}&+ 2\Big(\tfrac{\Delta - r + \frac{1}{2}}{\Delta - \frac{1}{2}}\Big) P\indices{^{{\color{red}(}a_1} _{c}}\bar{Q}^{(a_2{\color{red})}}\ket{\Delta, r}^{c)}\nonumber\\ &+ \Big(\tfrac{\Delta - r + \frac{1}{2}}{\Delta + \frac{1}{2}}\Big)P^{a_1 a_2}\bar{Q}_{c}\ket{\Delta,r}^{c}\nonumber \end{align}} \vspace{-1cm}\\
\hline
$[1]_{\Delta + \frac{3}{2}, r + 1}$& \vspace{-1cm}
{\begin{align}Q^{c}Q_{c}\bar{Q}^{(a_1}\ket{\Delta, r}^{a_2)}&- 2\Big(\tfrac{\Delta + r + \frac{1}{2}}{\Delta - \frac{1}{2}}\Big) P\indices{^{{\color{red}(}a_1} _c}Q^{(a_2{\color{red})}}\ket{\Delta, r}^{c)}\nonumber\\ &+ \Big(\tfrac{\Delta + r + \frac{1}{2}}{\Delta + \frac{1}{2}}\Big)P^{a_1 a_2}Q_{c}\ket{\Delta,r}^{c}\nonumber \end{align}} \vspace{-1cm}\\
\hline
\multicolumn{2}{|c|}{\textbf{Level 4} 
\footnote{Note that at level 4, naively there are two possible $P^2$ terms, namely $P^{bc}P_{bc}\ket{\Delta}^{a_1}$ and $P^{a_1}_{\ \ b}P^{b}_{\ \ c}\ket{\Delta}^{c}$.  However we have 
\begin{multline} P^{a_1}_{\ \ b}P^{b}_{\ \ c}\ket{\Delta}^{c}=P^iP^j \left( \sigma_i\sigma_j\right)^{a_1}_{\ \ c} \ket{\Delta}^{c}=P^iP^j \left({1\over 2} \left\{ \sigma_i,\sigma_j\right\}\right)^{a_1}_{\ \ c} \ket{\Delta}^{c} \\ =P^iP^j \delta_{ij} \left( {\mathbb I}\right)^{a_1}_{\ \ c} \ket{\Delta}^{c}=P^2\ket{\Delta}^{a_1}=-{1\over 2}P^{bc}P_{bc} \ket{\Delta}^{a_1}, \nonumber \end{multline}
thus there is really only one spin-1/2 $P^2$ descendent, see e.g. \cite{deWit:1999ui}.}
}\\
\hline
$[\frac{1}{2}]_{\Delta + 2, r}$ & \vspace{-1cm}
{\begin{align}&\bar{Q}^{c}\bar{Q}_{c}Q^{d}Q_{d}\ket{\Delta, r}^{a_1}
-2\tfrac{(\Delta - r - \frac{3}{2})(\Delta - r + \frac{1}{2})}{\Delta(\Delta - \frac{3}{2})}P^{cd}P_{cd}\ket{\Delta, r}^{a_1}\nonumber \\
&+ 4\Big(\tfrac{\Delta - r - \frac{3}{2}}{\Delta - \frac{3}{2}}\Big) P_{cd}\bar{Q}^{(c}Q^{d}\ket{\Delta,r}^{a_1)} - \tfrac{4}{3}\Big(\tfrac{\Delta - r +\frac{3}{2}}{\Delta}\Big)P\indices{^{a_1} _c}\bar{Q}^{d}Q_{d}\ket{\Delta,r}^{c} 
\nonumber\\
&-\tfrac{8}{3}\tfrac{(\Delta - r)}{\Delta} P\indices{^{a_1} _c}\bar{Q}_{d}Q^{c}\ket{\Delta,r}^{d}\nonumber \end{align}} \vspace{-1cm} \\
\hline
\end{longtable}
\end{center}

\newpage

\clearpage

 \subsection{$s\geq 1$}
\begin{figure}[h!]
\begin{center}
\epsfig{file=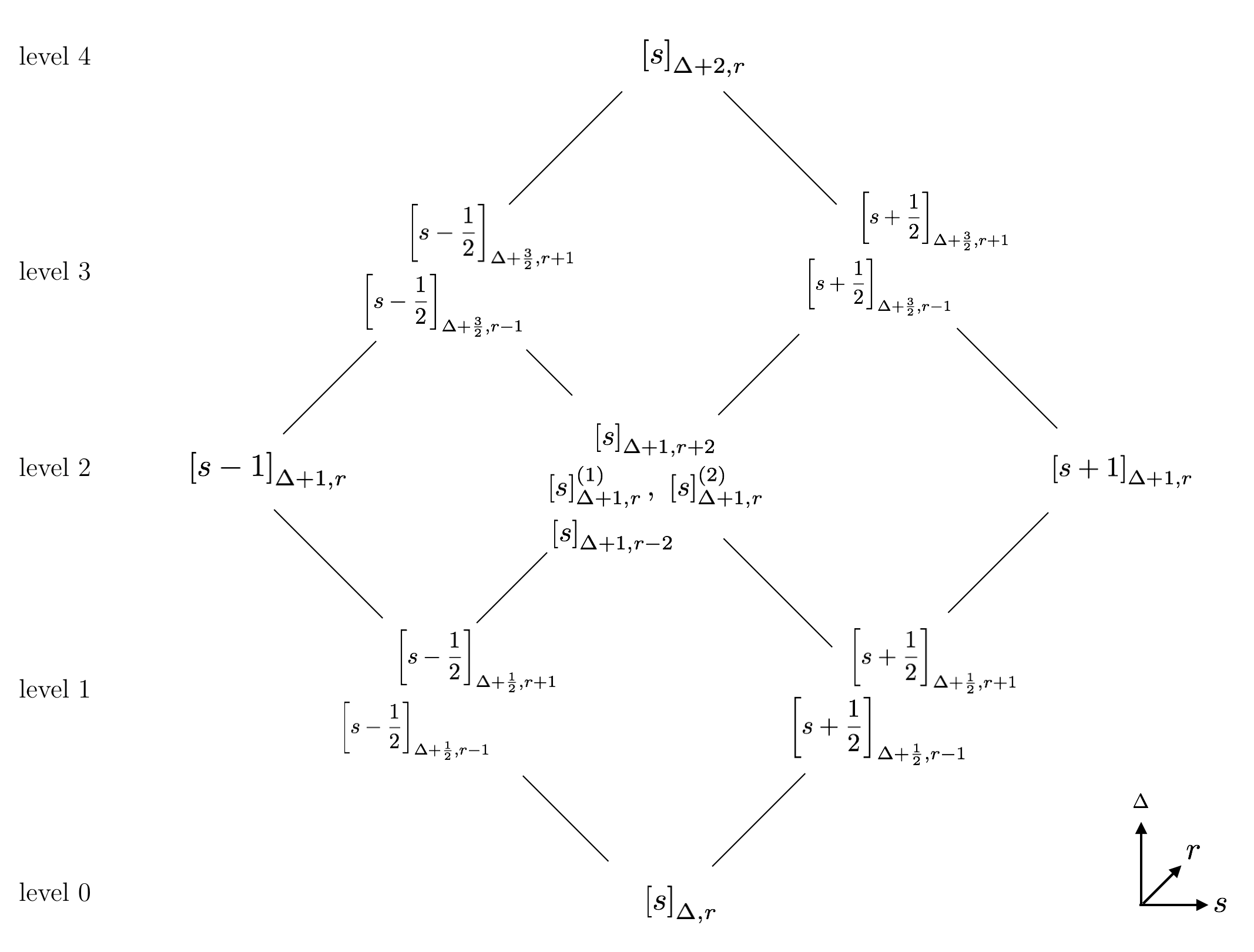,width=6.0in}
\caption{\small  Conformal primaries of the generic multiplet $\left\{ {s}\right\}_{\Delta,r}$.  The $[{s}]_{\Delta, k}$ conformal primary at level zero is also the superconformal primary.}
\label{s1multipletgeneric}
\end{center}
\end{figure}

The conformal primaries of the generic $\{s\}_{\Delta, r}$ multiplet with $s\geq 1$ are shown in Table \ref{sconformalprimary}.  Note that at level 2 there are two different states with spin $s$, weight $\Delta+1$, and charge $r$, these are denoted by $[s]^{(1)}_{\Delta + 1, r}$ and $[s]^{(2)}_{\Delta + 1, r}$.
This multiplet can be visualized as in Figure \ref{s1multipletgeneric}.  From the bulk point of view, for generic values of $\Delta$ this multiplet contains one massive spin $s-1$ field, 4 massive spin $s-{1\over 2}$ fields, 6 massive spin $s$ fields, 4 massive spin $s+{1\over 2}$ fields, and one massive spin $s+1$ field.  Using the fact that a massive spin $s$ field has $2s+1$ propagating degrees of freedom, the multiplet is seen to contain $8(2s+1)$ bosonic degrees of freedom and an equal number of fermionic degrees of freedom.

\bigskip
\begin{center}
\begin{longtable}{ |m{3.25 cm}| m{11.5 cm}| }
\caption{Conformal primaries in the generic $\{s\}_{\Delta, r}$ superconformal multiplet.}
\label{sconformalprimary}\\
\hline
\multicolumn{2}{|c|}{\textbf{Conformal Primaries of $\{s\}_{\Delta,r}$}}\\
\hline
\multicolumn{2}{|c|}{\textbf{Level 1}}\\
\hline
$[s - \frac{1}{2}]_{\Delta + \frac{1}{2}, r - 1}$ & $\bar{Q}_{b}\ket{\Delta, r}^{a_1...a_{2s-1}b}$\\
\hline
$[s - \frac{1}{2}]_{\Delta + \frac{1}{2}, r + 1}$ & $Q_{b}\ket{\Delta, r}^{a_1...a_{2s-1}b}$\\
\hline
$[s + \frac{1}{2}]_{\Delta + \frac{1}{2}, r - 1}$ & $\bar{Q}^{(a}\ket{\Delta, r}^{a_1...a_{2s})}$ \\
\hline
$[s + \frac{1}{2}]_{\Delta + \frac{1}{2}, r + 1}$ & $Q^{(a}\ket{\Delta, r}^{a_1...a_{2s})}$ \\
\hline
\multicolumn{2}{|c|}{\textbf{Level 2}}\\
\hline
$[s - 1]_{\Delta + 1, r}$& $\bar{Q}_{a}Q_{b}\ket{\Delta, r}^{a_1 ...a_{2s-2}ab}  - \Big(\frac{\Delta - r - s-1}{\Delta - s-1}\Big)P^{(a b}\ket{\Delta, r}^{a_1...a_{2s})}$\\
\hline
$[s]^{(1)}_{\Delta + 1, r}$ & $\bar{Q}^{c}Q_{c}\ket{\Delta, r}^{a_1...a_{2s}} + \Big(\frac{2s}{\Delta - 1}\Big)P\indices{^{(a_1}_b}\ket{\Delta, r}^{a_2...a_{2s})b}$ \\
\hline
$[s]^{(2)}_{\Delta + 1, r}$& $\bar{Q}_{b}Q^{(a_1}\ket{\Delta, r}^{a_2...a_{2s})b} - \Big(\frac{\Delta + s - r - 1}{\Delta - 1}\Big)P\indices{^{(a_1}_b}\ket{\Delta, r}^{a_2...a_{2s})b}$\\
\hline
$[s]_{\Delta + 1, r - 2}$ & $\bar{Q}^{c}\bar{Q}_{c}\ket{\Delta, r}^{a_1...a_{2s}}$\\
\hline
$[s]_{\Delta + 1, r + 2}$ & $Q^{c}Q_{c}\ket{\Delta, r}^{a_1...a_{2s}}$\\
\hline
$[s + 1]_{\Delta + 1, r}$& $\bar{Q}^{(a}Q^{b}\ket{\Delta, r}^{a_1...a_{2s})}  - \Big(\frac{\Delta - r + s}{\Delta + s}\Big)P^{(a b}\ket{\Delta, r}^{a_1...a_{2s})}$\\
\hline
\newpage
\hline
\multicolumn{2}{|c|}{\textbf{Level 3}}\\
\hline
$[s - \frac{1}{2}]_{\Delta + \frac{3}{2}, r - 1}$  & \vspace{-1cm}
 {\begin{align}&\bar{Q}^{c}\bar{Q}_{c}Q_{b}\ket{\Delta, r}^{a_1...a_{2s-1}b} \nonumber \\&-2\Big(\tfrac{1 - 2s}{1 + 2s}\Big)\Big(\tfrac{\Delta - r - s - 1}{\Delta - \frac{1}{2}}\Big) P\indices{^{(a_1} _b}\bar{Q}_{d}\ket{\Delta, r}^{a_2...a_{2s - 1})bd} \nonumber \\ &-2\Big(\tfrac{\Delta - r - s - 1}{\Delta - s - 1}\Big) P_{bd}\bar{Q}^{(a_1}\ket{\Delta, r}^{a_1...a_{2s-1}bd)} \nonumber \end{align}} \vspace{-1cm}\\
\hline
$[s - \frac{1}{2}]_{\Delta + \frac{3}{2}, r + 1}$  & \vspace{-1cm}
{\begin{align}&Q^{c}Q_{c}\bar{Q}_{b}\ket{\Delta, r}^{a_1...a_{2s-1}b} \nonumber \\&-2\Big(\tfrac{1 - 2s}{1 + 2s}\Big)\Big(\tfrac{\Delta + r - s - 1}{\Delta - \frac{1}{2}}\Big) P\indices{^{(a_1} _b}Q_{d}\ket{\Delta, r}^{a_2...a_{2s - 1})bd} \nonumber \\ &-2\Big(\tfrac{\Delta + r - s - 1}{\Delta - s - 1}\Big) P_{bd}Q^{(a_1}\ket{\Delta, r}^{a_1...a_{2s-1}bd)} \nonumber \end{align}} \vspace{-1cm}\\
\hline
$[s + \frac{1}{2}]_{\Delta + \frac{3}{2}, r - 1}$& \vspace{-1cm}
 {\begin{align}\bar{Q}^{c}\bar{Q}_{c}Q^{(a}\ket{\Delta, r}^{a_1...a_{2s})}&- 2\Big(\tfrac{\Delta - r + s}{\Delta - \frac{1}{2}}\Big) P\indices{^{{\color{red}(}a} _b}\bar{Q}^{(a_1}\ket{\Delta, r}^{a_2...a_{2s}{\color{red})}b)}\nonumber\\ &+ \tfrac{4s(\Delta - r + s)}{(2s + 1)(\Delta + s)}P^{(a a_1}\bar{Q}_{b}\ket{\Delta,r}^{a_2...a_{2s})b}\nonumber \end{align}} \vspace{-1cm}\\
\hline
$[s + \frac{1}{2}]_{\Delta + \frac{3}{2}, r + 1}$& \vspace{-1cm}
{\begin{align}Q^{c}Q_{c}\bar{Q}^{(a}\ket{\Delta, r}^{a_1...a_{2s})}&- 2\Big(\tfrac{\Delta + r + s}{\Delta - \frac{1}{2}}\Big) P\indices{^{{\color{red}(}a} _b}Q^{(a_1}\ket{\Delta, r}^{a_2...a_{2s}{\color{red})}b)}\nonumber\\ &+ \tfrac{4s(\Delta + r + s)}{(2s + 1)(\Delta + s)}P^{(a a_1}Q_{b}\ket{\Delta,r}^{a_2...a_{2s})b}\nonumber \end{align}} \vspace{-1cm}\\
\hline
\multicolumn{2}{|c|}{\textbf{Level 4}}\\
\hline
$[s]_{\Delta + 2, r}$ & \vspace{-1cm}
{\begin{align}&\bar{Q}^{c}\bar{Q}_{c}Q^{d}Q_{d}\ket{\Delta, r}^{a_1...a_{2s}}\nonumber \\
&-2\Big(\tfrac{1 + 2 s}{1 + s}\Big)\tfrac{\Delta(\Delta - r - s - 1)(\Delta-r + s)}{(\Delta - s - 1)(\Delta+s)(\Delta - \frac{1}{2})} P_{ab}P^{(a b}\ket{\Delta,r}^{a_1 a_2...a_{2s})}\nonumber \\&+4\tfrac{s^2}{(1 + s)^2}\tfrac{(\Delta - r - s - 1)(\Delta + s + 1)(\Delta - r + s)}{\Delta(\Delta+s)(\Delta - \frac{1}{2})} P\indices{^{{\color{red}(}a_1} _c}P\indices{^{(a_2} _b}\ket{\Delta,r}^{a_3...a_{2s}{\color{red})}c) b}\nonumber \\
&+ 4 \Big(\tfrac{2s - 1}{2s + 1}\Big)\Big(\tfrac{\Delta - r + s}{\Delta+s}\Big) P^{(a_1 a_2}\bar{Q}_{c}Q_{d}\ket{\Delta,r}^{a_3...a_{2s})cd}\nonumber \\
&+ 4\Big(\tfrac{\Delta - r - s - 1}{\Delta - s - 1}\Big) P_{ab}\bar{Q}^{(a}Q^{b}\ket{\Delta,r}^{a_1 a_2...a_{2s})}\nonumber \\& - 4\Big(\tfrac{s}{s + 1}\Big)\Big(\tfrac{\Delta - r + s + 1}{\Delta}\Big) \epsilon^{d c}P\indices{^{(a_1} _b}\bar{Q}_{d}Q_{c}\ket{\Delta,r}^{a_2...a_{2s}) b} \nonumber \\&-8\Big(\tfrac{s}{s + 1}\Big)\tfrac{(\Delta - r)}{\Delta} P\indices{^{{\color{red}(}a_1} _c}\bar{Q}_{b}Q^{(a_2}\ket{\Delta,r}^{a_3...a_{2s}{\color{red})}c) b}\nonumber \end{align}}  \vspace{-1cm} \\
\hline
\end{longtable}
\end{center}

\newpage

\clearpage

\section{Shortening conditions}

The superconformal primary is taken to be normalized to unity according to 
\begin{equation}
    \,_{b_1...b_{2s}}\braket{\Delta, r|\Delta, r}^{a_1...a_{2s}} = \delta\indices{^{(a_1}_{b_1}}...\delta\indices{^{a_{2s})}_{b_{2s}}},
\end{equation}
where we have introduced the conjugate $\ket{\Delta, r}^{a_1...a_{2s}\dagger} = \,_{a_1...a_{2s}}\bra{\Delta, r}$.  Given this, we can then compute the norms of the remaining conformal primaries by repeatedly applying the superconformal algebra (anti)commutation rules in Section \eqref{susycalgesecef}.  At a given $s,r$, for large enough $\Delta$ all the norms will be positive, and the representation is unitary.  As we lower $\Delta$, some norms may pass through zero and then become negative.  

The superconformal multiplet shortens for certain values of $\Delta$ where a conformal primary becomes null, meaning it has zero norm and zero overlap with all other states in the superconformal multiplet.  At these values, the null states can be factored out, leaving a shorter multiplet.  These short multiplets are unitary only if the remaining norms are all positive.  

In this section, we compute the norms of the conformal primaries and the values of $\Delta$ at which these shortenings occur, and study the structure of the some of the resulting short representations.  $s = 0$ and $s = \frac{1}{2}$ are exceptional cases, since certain conformal primaries do not exist within the superconformal multiplet, so we treat these separately in the sections that follow.   

\subsection{$s=0$\label{s0normsubsec}}

\begin{figure}
\begin{center}
\epsfig{file=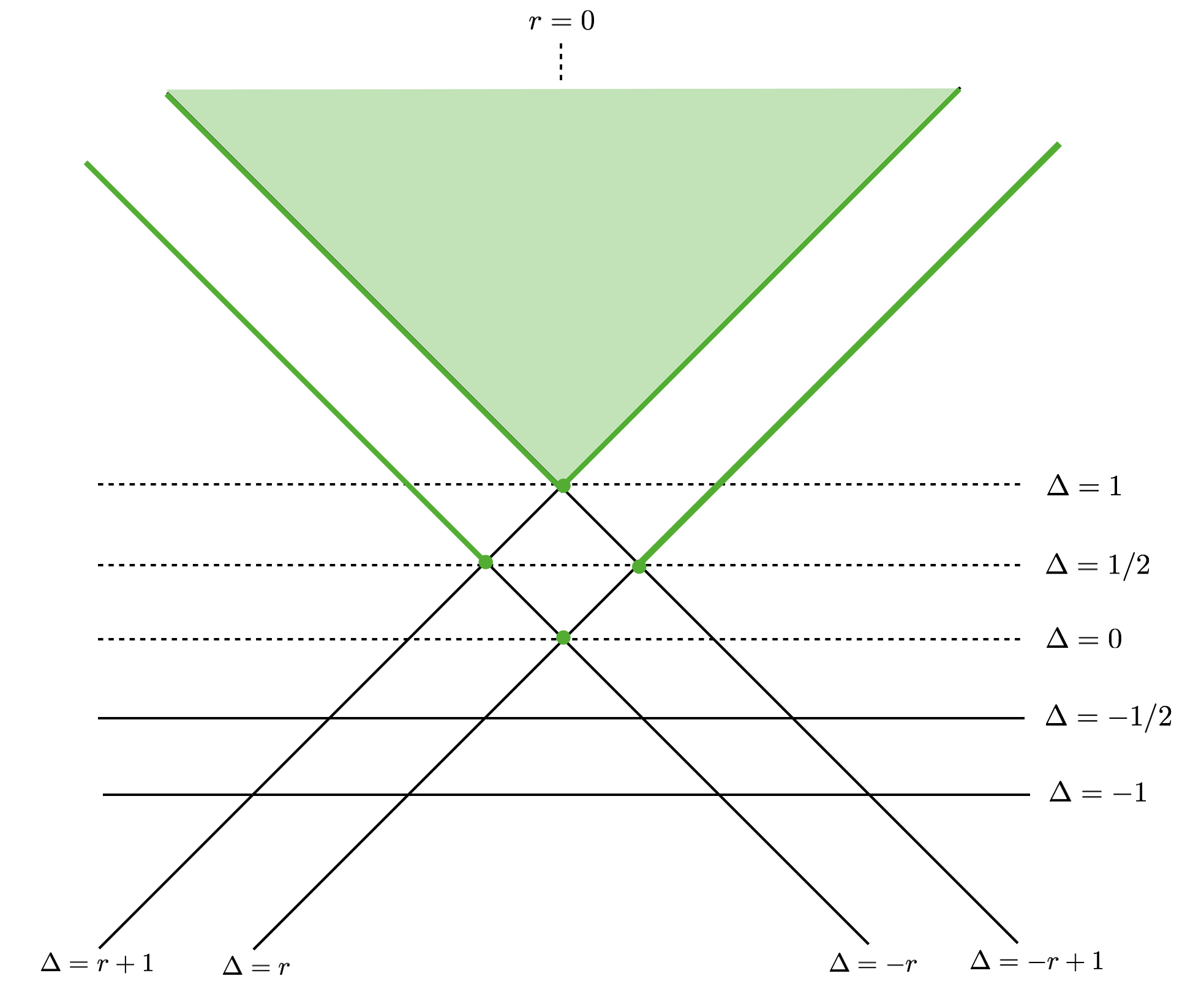,width=5.0in}
\caption{\small  Shortening conditions for the $\{0\}_{\Delta,r}$ superconformal multiplet as they occur in the $r,\Delta$ plane.  Solid lines are places where states become null and the multiplet shortens.  Dotted lines are places where the extended modules  discussed in Section \ref{extendedmodulesection} occur.  Regions in green are unitary, where all non-null norms are positive. All other regions are non-unitary, meaning at least one non-null norm is negative.}
\label{shortsplane0}
\end{center}
\end{figure}

\begin{figure}
\begin{center}
\epsfig{file=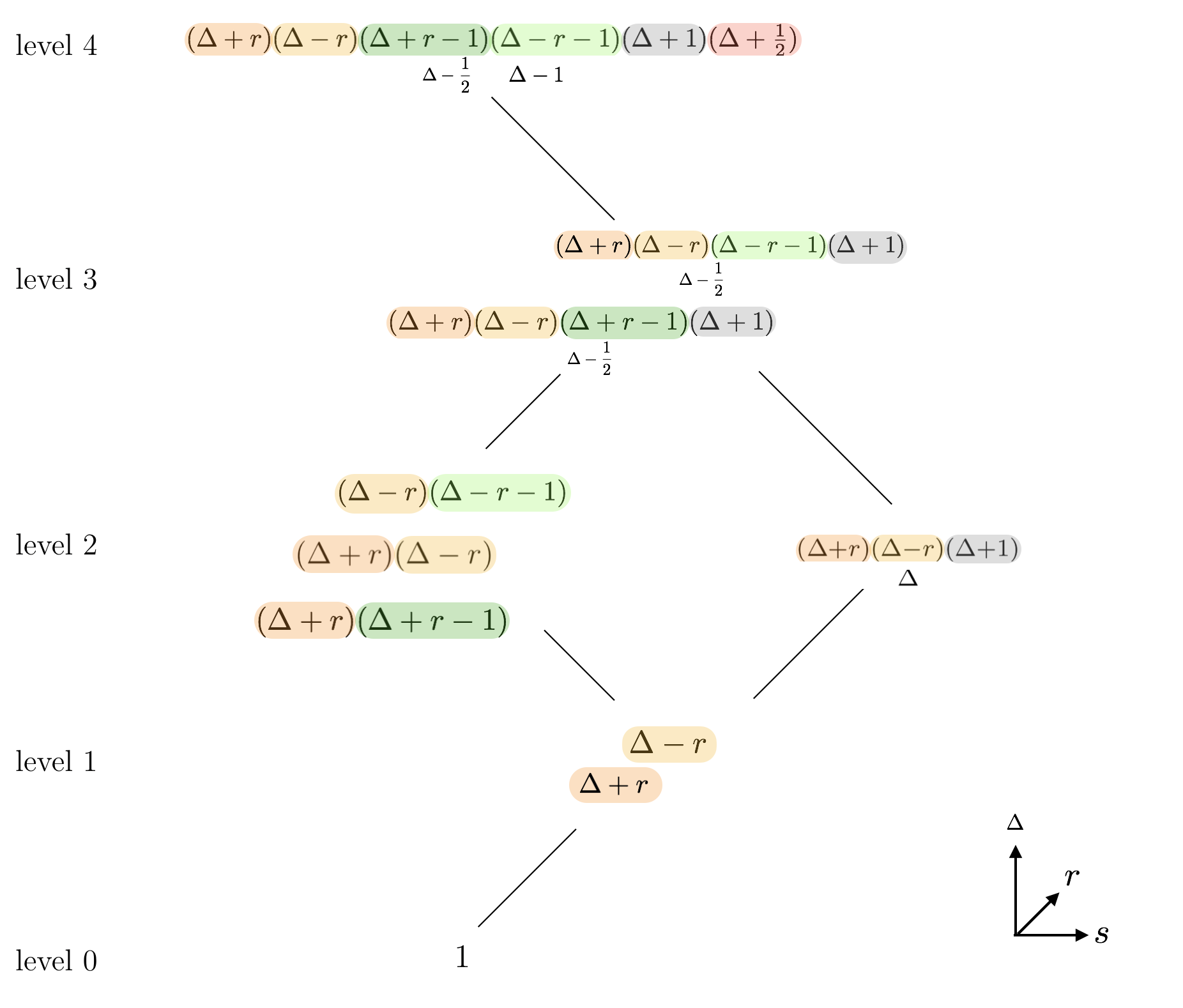,width=5.0in}
\caption{\small Shortening conditions at each level for the $\{0\}_{\Delta,r}$ superconformal multiplet.  The placement of the states corresponds to the states in Figure \ref{s0multipletgeneric}.  The colored factors are factors in the numerators of the norms in Table \ref{s0norms}.  Those with the same color vanish at the same value of $\Delta$, where the states go null and shortening occurs.  The uncolored factors underneath are values where primaries in Table \ref{s0conformalprimary} become singular.  At these values the extended modules discussed in Section \ref{extendedmodulesection} occur.}
\label{shortsmultiplet0}
\end{center}
\end{figure}

For the $\{0\}_{\Delta, r}$ superconformal multiplet, the norm of each conformal primary in Table \ref{s0conformalprimary} is tabulated in Table \ref{s0norms}.  Shortening conditions occur when these norms vanish.  The norms can also become singular for certain values of $\Delta$.  When this happens, one of the denominators of the $P$ terms in the conformal primary is becoming singular.   When the denominator of one of the $P$ terms is singular, we have instead the phenomena of extended modules, as discussed in Section \ref{extendedmodulesection}.  The shortening and extended module values as they occur in the $r,\Delta$ plane are visualized in Figure \ref{shortsplane0}, and as they occur in the various states of the multiplet in Figure \ref{shortsmultiplet0}.

\newpage
\begin{center}
\begin{longtable}{|m{5cm}| m{9cm}|}
\caption{Norms of the conformal primaries in the $\{0\}_{\Delta, r}$ superconformal multiplet, as they are shown in Table \ref{s0conformalprimary}. The norms also include totally symmetric products of $\delta$-functions in spinor indices which we have omitted for brevity.}
\label{s0norms}\\
\hline
\multicolumn{2}{|c|}{\textbf{Conformal Primary Norms: $\{0\}_{\Delta,r}$}}\\
\hline
{\textbf{Conformal Primary}} & {\textbf{Norm}}\\
\hline
\multicolumn{2}{|c|}{\textbf{Level 1}}\\
\hline
$\left[ \frac{1}{2}\right]_{\Delta + \frac{1}{2},r - 1}$ & $2(\Delta +  r)$\\
\hline
 $\left[ \frac{1}{2}\right]_{\Delta + \frac{1}{2},r + 1}$ & $2(\Delta - r)$\\
\hline
\multicolumn{2}{|c|}{\textbf{Level 2}}\\
\hline
  $\left[0\right]_{\Delta +1,r }$ & $8(\Delta + r)(\Delta - r)$\\
\hline
 $\left[0\right]_{\Delta +1,r - 2}$  & $16(\Delta  + r)(\Delta  + r - 1)$ \\
\hline
 $\left[0\right]_{\Delta +1,r + 2}$  & $16(\Delta - r)(\Delta - r - 1)$ \\
\hline
 $\left[1\right]_{\Delta +1,r }$ & $4\frac{(\Delta + r)(\Delta - r)(\Delta + 1)}{\Delta}$ \\
\hline
\multicolumn{2}{|c|}{\textbf{Level 3}}\\
\hline
 $\left[ \frac{1}{2}\right]_{\Delta + \frac{3}{2},r - 1}$ & $32\frac{(\Delta + r)(\Delta - r)(\Delta + r - 1)(\Delta + 1)}{\Delta - {1\over 2}}$\\
\hline
 $\left[ \frac{1}{2}\right]_{\Delta + \frac{3}{2},r + 1}$ & $32\frac{(\Delta + r)(\Delta - r)(\Delta - r - 1)(\Delta + 1)}{\Delta - {1\over 2}}$\\
\hline
\multicolumn{2}{|c|}{\textbf{Level 4}}\\
\hline
 $\left[0\right]_{\Delta +2,r }$ & $256\frac{(\Delta + r)(\Delta - r)(\Delta + r - 1)(\Delta - r - 1)(\Delta + 1)(\Delta + \frac{1}{2})}{(\Delta - 1)(\Delta - \frac{1}{2})}$\\
\hline
\end{longtable}
\end{center}

\newpage

\clearpage

\newpage

\clearpage

\newpage
\begin{figure}
\begin{center}
\epsfig{file=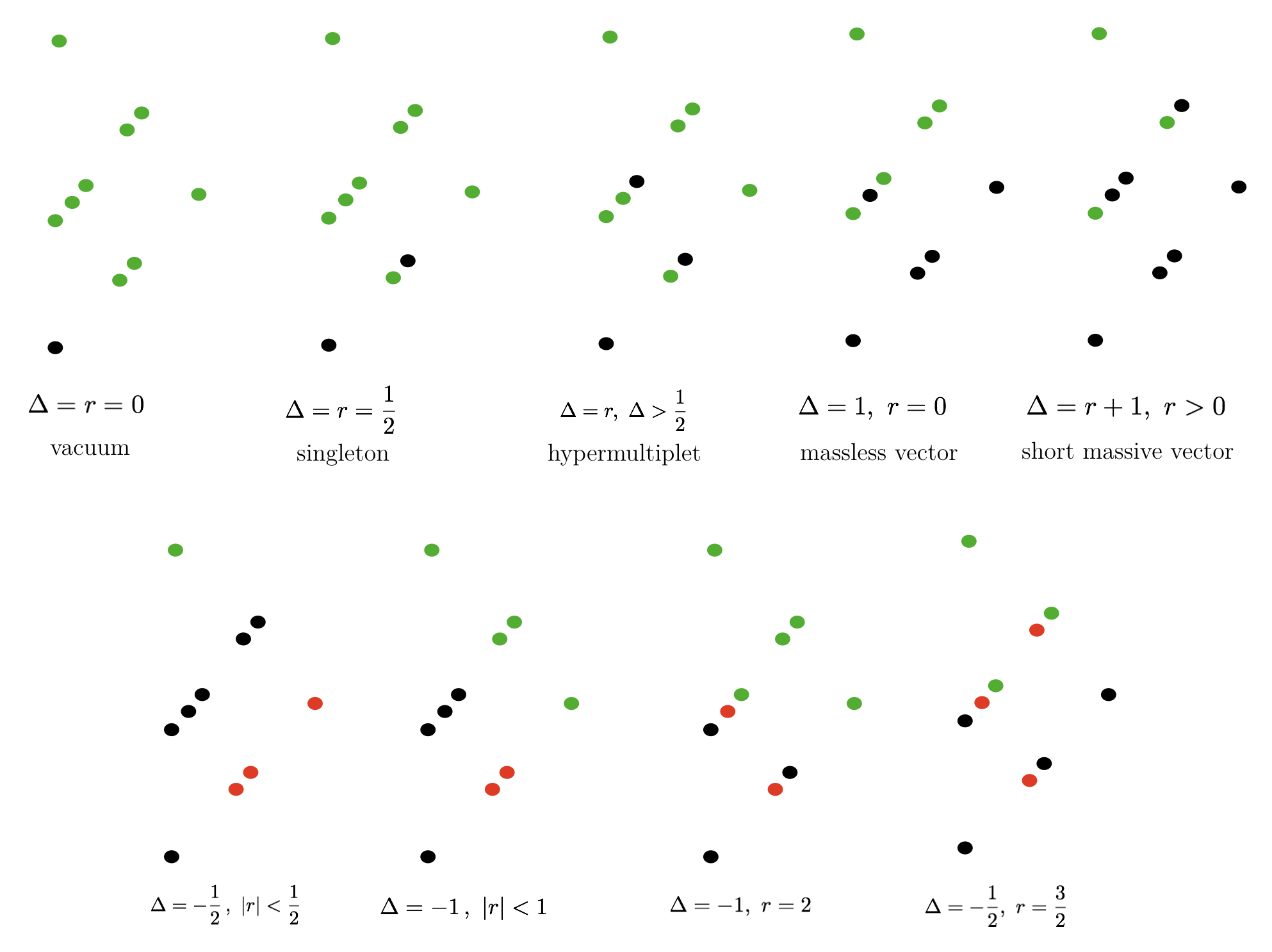,width=6.0in}
\caption{\small Short $\{0\}_{\Delta,r}$ multiplets.  Black states have positive norm, red states have negative norm, and green states are the zero norm null states that decouple from the multiplet.  The placement of the states corresponds to the states in Figure \ref{s0multipletgeneric}.  The first line contains all the different unitary short multiplets.  The second line contains some of the new non-unitary short multiplets.}
\label{individualshortsmultiplet0}
\end{center}
\end{figure}

\clearpage

In Figure \ref{individualshortsmultiplet0} we show some of the shortened multiplets.  The known unitary short multiplets are as follows: there is the vacuum multiplet at $\Delta=r=0$, where all the higher states are null.  This is an isolated unitary multiplet \cite{Cordova:2016emh}, as can be seen from Figure \ref{shortsplane0}.   Along the green lines at $\Delta=\pm r$, we have the scalar hypermultiplets which consist of two massive scalars are a massive spin 1/2 (these are the standard ``matter'' multiplets in ${\cal N}=2$ SUSY field theories in $D=4$).  At the endpoints of these lines, at $\Delta=\pm r=1/2$, we have a further shortening that removes a scalar and leaves the supersymmetric singleton \cite{Dirac:1963ta,Gunaydin:1989um,Flato:1999yp,Bekaert:2011js}.  These fields have no bulk propagating degrees of freedom.  From the point of view of CFT it is a free scalar and free fermion\footnote{Note that the singleton does not split upon reduction to ${\cal N}=1$, it also appears as an ${\cal N}=1$ multiplet \cite{Garcia-Saenz:2018wnw},
\be \left\{0\right\}_{{1\over 2}}^{{\cal N}=2}=\left\{0\right\}_{{1\over 2}}^{{\cal N}=1} .\ee}.   Along the lines $\Delta=\pm r+1$, we have a short massive spin-1 multiplet, consisting of a massive spin-1, 3 massive spin 1/2's, and 3 scalars, for a total of 6 bosonic and 6 fermionic bulk degrees of freedom.  (As discussed in Section \ref{branchingsection}, this multiplet appears in the branching rules of a massive spin 3/2 multiplet as it approaches its massless value.)  Where these lines meet, at $\Delta=1$, $r=0$ we have a further shortening to the massless photon multiplet consisting of a massless spin 1, two scalars, and two spin 1/2's, for a total of 4 bosonic and 4 fermionic degrees of freedom (this is the standard ``gauge'' multiplet of ${\cal N}=2$ SUSY field theories in $D=4$).  These unitary multiplets are all shown in the first line of Figure \ref{individualshortsmultiplet0}.  In addition to these known unitary short multiplets, we find new non-unitary short multiplets along the solid black lines in Figure \ref{shortsplane0}.  Some of these are shown in the second line of Figure \ref{individualshortsmultiplet0}.


\subsection{$s=\tfrac{1}{2}$\label{s1/2normsubsec}}

For the $\{{1\over 2}\}_{\Delta, r}$ superconformal multiplet, the norm of each conformal primary in Table \ref{s1/2conformalprimary} is tabulated in Table  \ref{s1/2norms}.  Given that there are two different states with spin $1/2$, weight $\Delta+1$, and charge $r$, (the states $[\frac{1}{2}]^{(1)}_{\Delta + 1, r}$ and $[\frac{1}{2}]^{(2)}_{\Delta + 1, r}$ in Table \ref{s1/2conformalprimary}) there is a $2\times 2$ Graham matrix of inner products of these states.  This Graham matrix and its determinant are also shown. 

\begin{figure}
\begin{center}
\epsfig{file=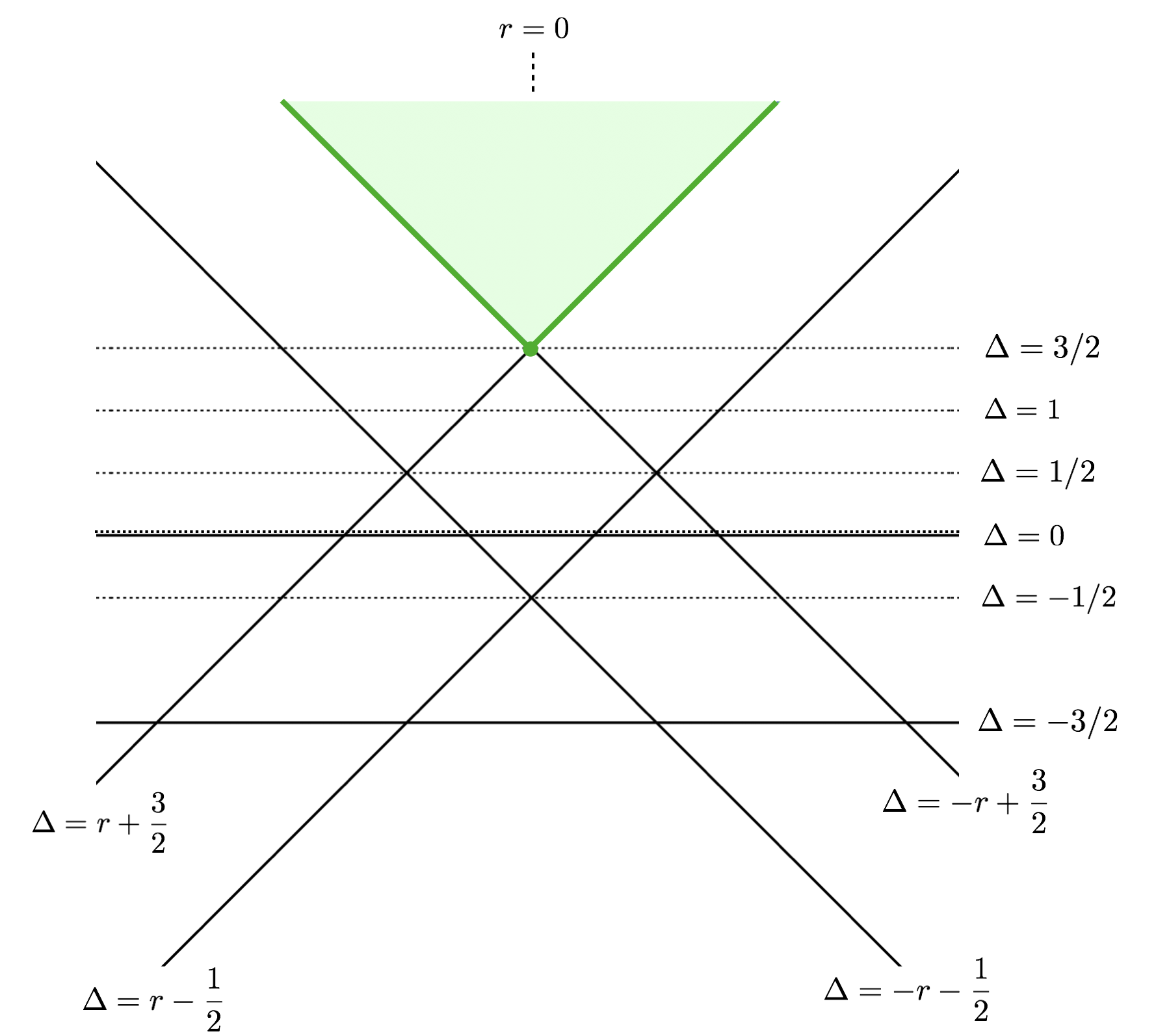,width=6.0in}
\caption{\small  Shortening conditions for the $\{{1\over 2}\}_{\Delta,r}$ superconformal multiplet as they occur in the $r,\Delta$ plane.  Solid lines are places where states become null and the multiplet shortens.  Dotted lines are places where the extended modules  discussed in Section \ref{extendedmodulesection} occur.  Regions in green are unitary, where all non-null norms are positive. All other regions are non-unitary, meaning at least one non-null norm is negative.}
\label{shortsplane1half}
\end{center}
\end{figure}

\begin{figure}
\begin{center}
\epsfig{file=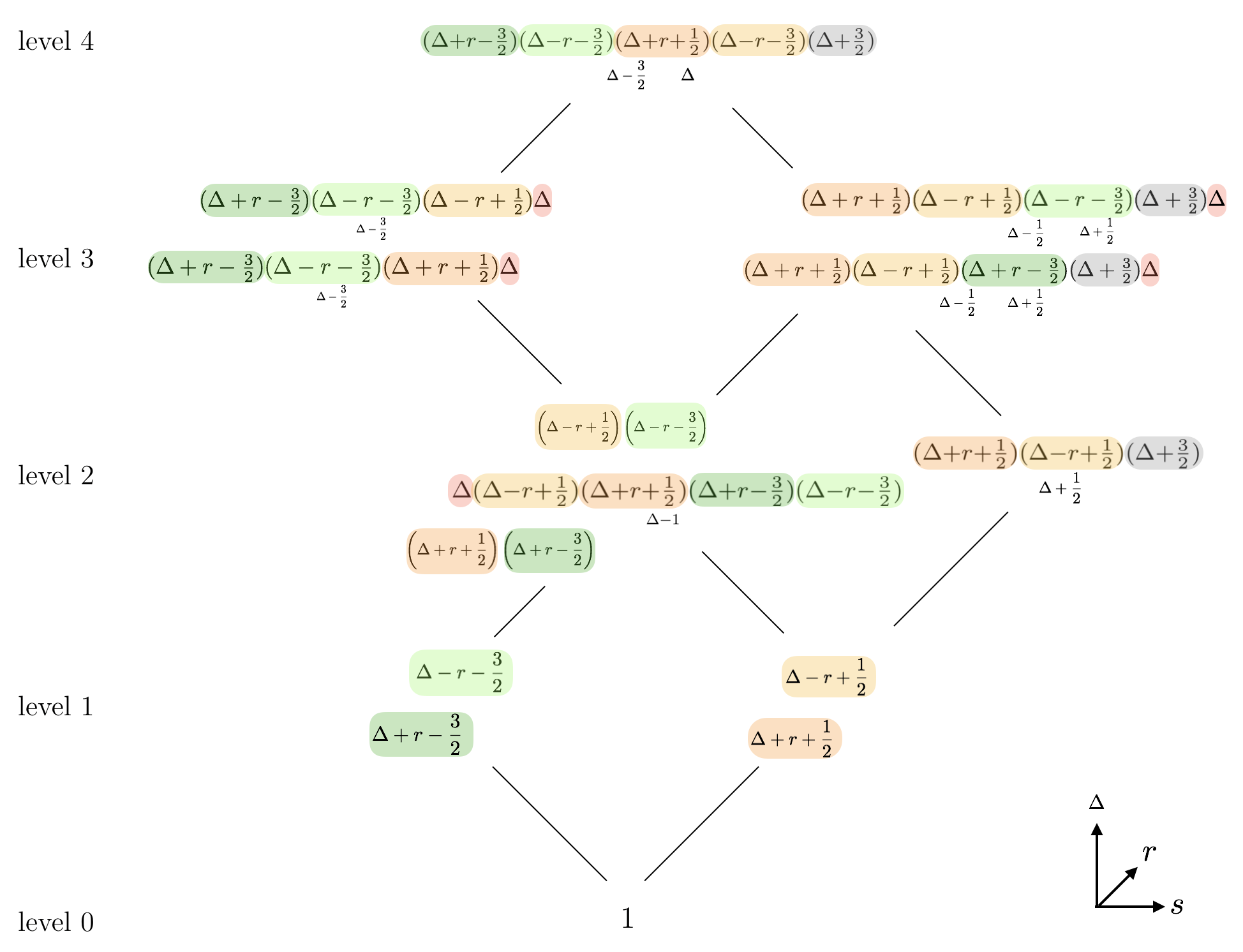,width=5.0in}
\caption{\small Shortening conditions at each level for the $\{{1\over 2}\}_{\Delta,r}$ superconformal multiplet.  The placement of the states corresponds to the states in Figure \ref{s1halfmultipletgeneric}.  The colored factors are factors in the numerators of the norms in Table \ref{s1/2norms}.  Those with the same color vanish at the same value of $\Delta$, where the states go null and shortening occurs.  The uncolored factors underneath are values where primaries in Table \ref{s1/2conformalprimary} become singular.  At these values the extended modules discussed in Section \ref{extendedmodulesection} occur.}
\label{shortsmultiplet1half}
\end{center}
\end{figure}

Shortenings and extended modules as they occur in the $r,\Delta$ plane are visualized in Figure \ref{shortsplane1half}, and as they occur in the various levels of the multiplet in Figure \ref{shortsmultiplet1half}.

\begin{center}
\begin{longtable}{ |m{5cm}| m{9cm}|}
\caption{Norms of the conformal primaries in the $\{\frac{1}{2}\}_{\Delta, r}$ superconformal multiplet, as they are shown in Table \ref{s1/2conformalprimary}. We have omitted totally symmetric products of $\delta$-functions in spinor indices for notational brevity.  
}
\label{s1/2norms}\\
\hline
\multicolumn{2}{|c|}{\textbf{Conformal Primary Norms: $\{\frac{1}{2}\}_{\Delta,r}$}}\\
\hline
{\textbf{Conformal Primary}} & {\textbf{Norm}}\\
\hline
\multicolumn{2}{|c|}{\textbf{Level 1}}\\
\hline
$[0]_{\Delta + \frac{1}{2}, r - 1}$ & $4(\Delta +  r - \frac{3}{2})$\\
\hline
$[0]_{\Delta + \frac{1}{2}, r + 1}$ & $4(\Delta - r - \frac{3}{2})$\\
\hline
$[1]_{\Delta + \frac{1}{2}, r - 1}$ & $2(\Delta +  r + \frac{1}{2})$\\
\hline
$[1]_{\Delta + \frac{1}{2}, r + 1}$ & $2(\Delta -  r + \frac{1}{2})$\\
\hline
\multicolumn{2}{|c|}{\textbf{Level 2}}\\
\hline
\multicolumn{2}{|c|}{$[\frac{1}{2}]^{(1,2)}_{\Delta + 1, r}$\,\,\,\, Graham Matrix  }\\
\hline
\multicolumn{2}{|c|}{$
\left(\begin{array}{cc} \frac{8(\Delta + r)(\Delta-r)(\Delta - 1) - 6\Delta}{\Delta - 1} & \frac{6r(1 - 2\Delta) + 4r^2(\Delta - 1) + \Delta(3 + 4\Delta - 4\Delta^2)}{\Delta - 1} \\  \frac{6r(1 - 2\Delta) + 4r^2(\Delta - 1) + \Delta(3 + 4\Delta - 4\Delta^2)}{\Delta - 1} & \frac{9 + 4r^2(1 - 4\Delta) - 28\Delta^2 + 16\Delta^3 + 12 r(2\Delta - 1)}{2(\Delta - 1)} \end{array}\right)$}\\
\hline $[\frac{1}{2}]^{(1,2)}_{\Delta + 1, r}$\,\,\,\, Graham Det. & $48\frac{\Delta(\Delta - r + \frac{1}{2})(\Delta + r + \frac{1}{2})(\Delta + r - \frac{3}{2})(\Delta - r - \frac{3}{2})}{\Delta - 1}$\\
\hline
$[\frac{1}{2}]_{\Delta + 1, r - 2}$ & $16(\Delta + r + \frac{1}{2})(\Delta + r - \frac{3}{2})$ \\
\hline
$[\frac{1}{2}]_{\Delta + 1, r + 2}$ & $16(\Delta - r + \frac{1}{2})(\Delta - r - \frac{3}{2})$ \\
\hline
$[\frac{3}{2}]_{\Delta + 1, r}$ & $4\frac{(\Delta + r + \frac{1}{2})(\Delta - r + \frac{1}{2})(\Delta + \frac{3}{2})}{\Delta + \frac{1}{2}}$ \\
\hline
\multicolumn{2}{|c|}{\textbf{Level 3}}\\
\hline
$[0]_{\Delta + \frac{3}{2}, r - 1}$& $64\frac{(\Delta + r - \frac{3}{2})(\Delta - r - \frac{3}{2})(\Delta +  r +\frac{1}{2})\Delta}{\Delta - \frac{3}{2}}$\\
\hline
$[0]_{\Delta + \frac{3}{2}, r + 1}$ & $64\frac{(\Delta + r - \frac{3}{2})(\Delta - r - \frac{3}{2})(\Delta - r +\frac{1}{2})\Delta}{\Delta - \frac{3}{2}}$\\
\hline
$[1]_{\Delta + \frac{3}{2}, r - 1}$ & $32\frac{(\Delta + r + \frac{1}{2})(\Delta - r + \frac{1}{2})(\Delta + r - \frac{3}{2})(\Delta + \frac{3}{2})\Delta}{(\Delta - \frac{1}{2})(\Delta + \frac{1}{2})}$\\
\hline
$[1]_{\Delta + \frac{3}{2}, r + 1}$ & $32\frac{(\Delta + r + \frac{1}{2})(\Delta - r + \frac{1}{2})(\Delta- r - \frac{3}{2})(\Delta + \frac{3}{2})\Delta}{(\Delta - \frac{1}{2})(\Delta + \frac{1}{2})}$\\
\hline
\multicolumn{2}{|c|}{\textbf{Level 4}}\\
\hline
$[\frac{1}{2}]_{\Delta + 2, r}$ & $256\frac{(\Delta + r - \frac{3}{2})(\Delta - r - \frac{3}{2})(\Delta + r + \frac{1}{2})(\Delta - r + \frac{1}{2})(\Delta + \frac{3}{2})}{(\Delta - \frac{3}{2})}$\\
\hline
\end{longtable}
\end{center}

\clearpage

The only unitary short multiplets are the short massive multiplets that occur along the lines $\Delta=3/2\pm r$ at the boundary of the green unitary region in Figure \ref{shortsplane1half}. They describe short multiplets for a massive spin $3/2$ particle whose structure is shown here.  In this picture and those like it below, black states have positive norm, red states have negative norm, and green states are zero norm null states that decouple from the multiplet.   For the degenerate state at level 2, the colors correspond to the two eigenvalues of the Graham matrix. The placement of the states is in accord with Figure \ref{s1halfmultipletgeneric}:
\be
\begin{array}{l} \left\{ {1\over 2} \right\}_{r+{3\over 2},r},\ |r|>0,\  \\ {\rm spin\ }  3/2  {\rm\ short\ massive\ multiplet:} \end{array}\ \ \ \ \ \ \ \ \raisebox{-96pt}{\epsfig{file=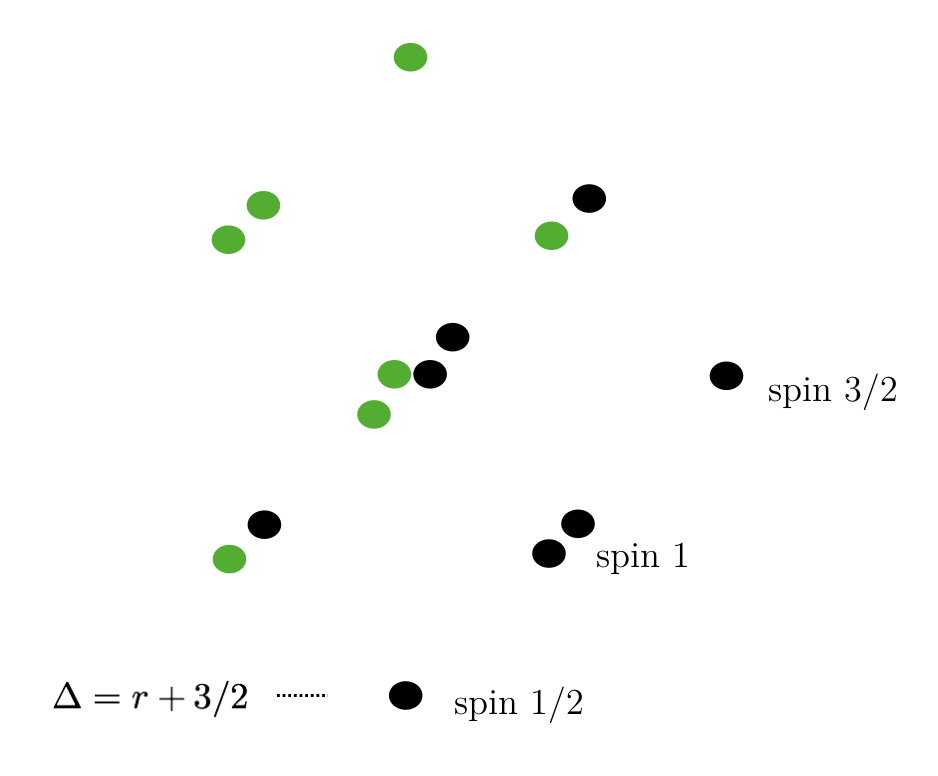,height=2.4in}}\label{massivemultipletpicse2}
\ee
This contains half as many degrees of freedom as the generic long massive multiplet.  As discussed in Section \ref{branchingsection}, this multiplet appears in the branching rules of a massive spin $2$ multiplet as it approaches its massless value.
When $r=0$, at the intersection of the two lines $\Delta=3/2\pm r$ at the apex of the triangular green unitary region in Figure \ref{shortsplane1half}, there is a further degeneration and we have the massless spin $3/2$ multiplet, shown in \eqref{PMmultipletpicse2}.  In the non-unitary region there are novel shortening conditions.  For example, at the intersection of the $\Delta=-{3\over 2}$ and $\Delta=-r-{1\over 2}$ shortening lines of Figure \ref{shortsplane1half}, we get the following short multiplet which only has spins $\leq 1$,
\be
\begin{array}{l} \left\{ {1\over 2} \right\}_{-{3\over 2},1}, \end{array}\ \ \ \ \ \ \ \ \raisebox{-96pt}{\epsfig{file=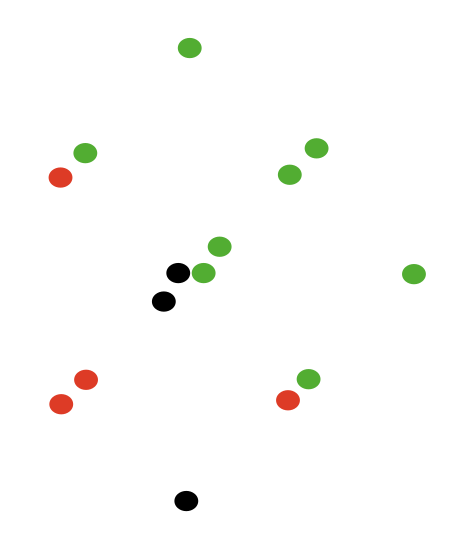,height=2.4in}}\label{massivemultipletpicse2}
\ee

\bigskip 

\subsection{$s\geq 1$\label{s1normsubsec}}

For the $\{s\}_{\Delta, r}$ superconformal multiplet with $s\geq 1$, the norm of each conformal primary in Table \ref{sconformalprimary}, up to an overall positive constant, is tabulated in Table  \ref{GenericNorms}.  Here we have used an extrapolation process to arrive at these norms for arbitrary $s$, which only gives us the values up to an overall positive constant.   Details about this extrapolation are presented in Appendix \ref{normsappendex}.  Given that there are two different states with spin $s$, weight $\Delta+1$, and charge $r$, (the states $[s]^{(1)}_{\Delta + 1, r}$ and $[s]^{(2)}_{\Delta + 1, r}$ in Table \ref{sconformalprimary}) there is a $2\times 2$ Graham matrix of inner products of these states.  This Graham matrix and its determinant are also shown.

\begin{figure}[h!]
\begin{center}
\epsfig{file=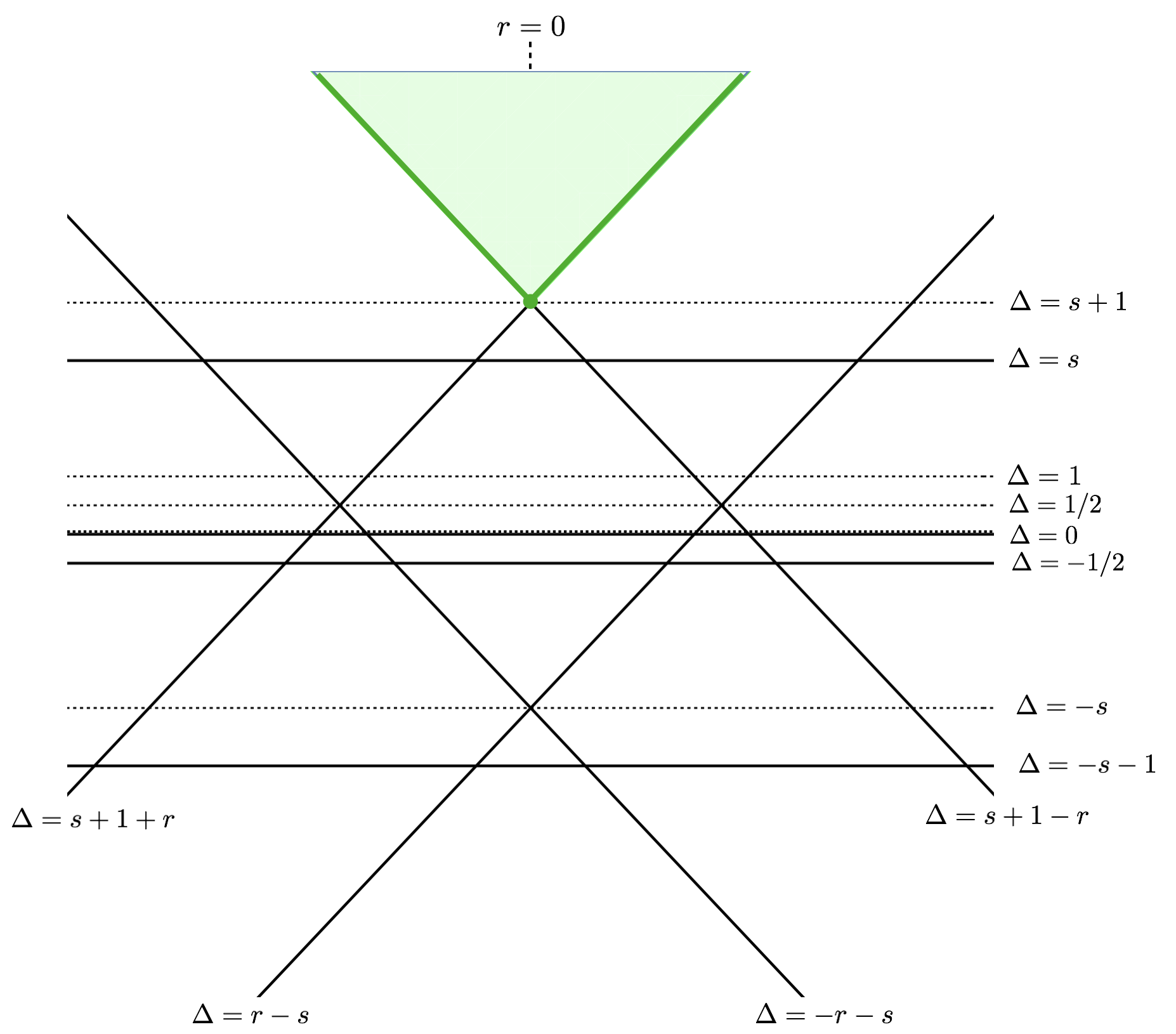,width=6.0in}
\caption{\small  Shortening conditions for the $\{{s}\}_{\Delta,r}$, $s\geq 1$ superconformal multiplet as they occur in the $r,\Delta$ plane.  Solid lines are places where states become null and the multiplet shortens.  Dotted lines are places where the extended modules  discussed in Section \ref{extendedmodulesection} occur.  Regions in green are unitary, where all non-null norms are positive. All other regions are non-unitary, meaning at least one non-null norm is negative.}
\label{shortsplanes}
\end{center}
\end{figure}

\begin{figure}[h!]
\begin{center}
\epsfig{file=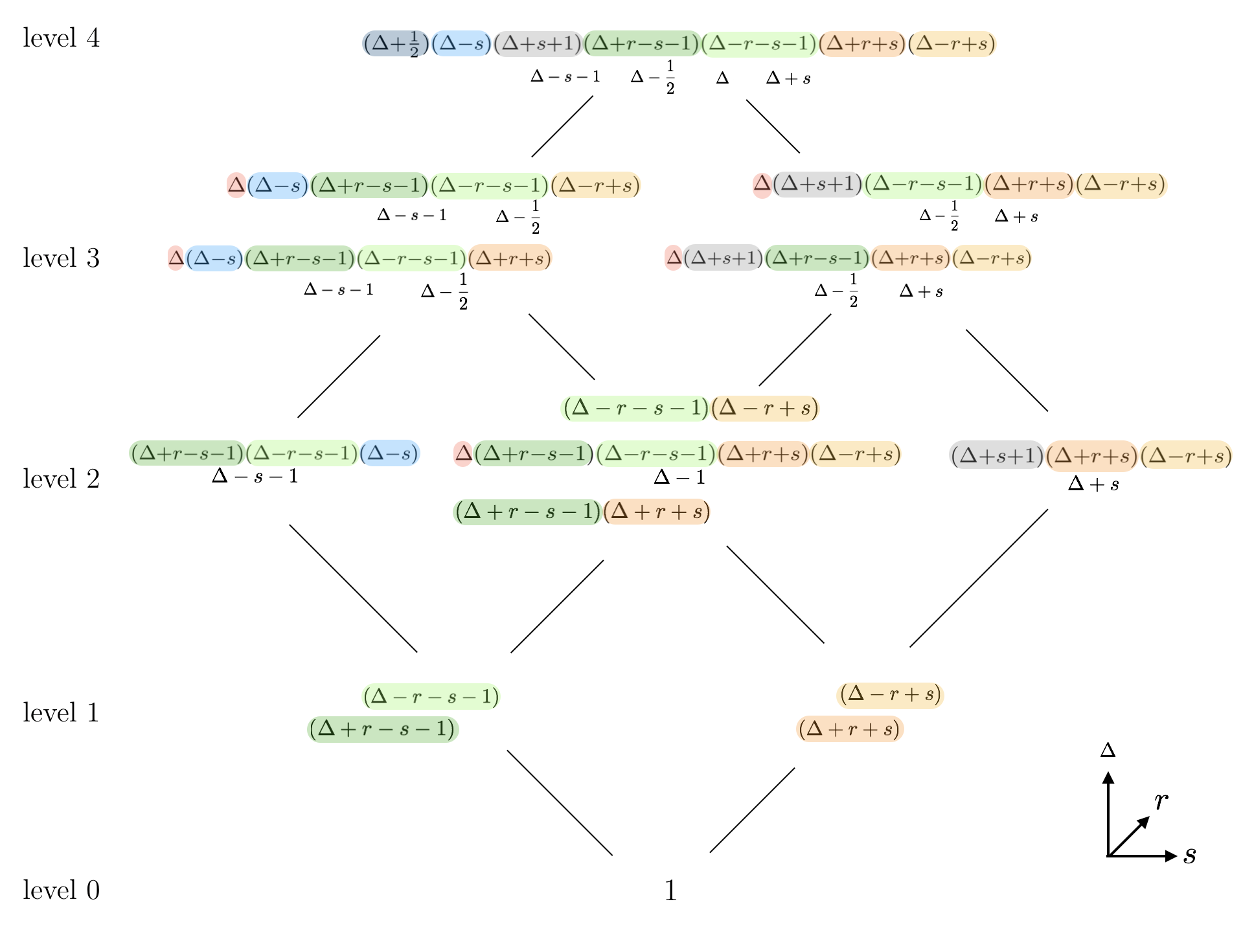,width=6.0in}
\caption{\small Shortening conditions at each level for the $\{{s}\}_{\Delta,r}$, $s\geq 1$ superconformal multiplet.  The placement of the states corresponds to the states in Figure \ref{s1multipletgeneric}.  The colored factors are factors in the numerators of the norms in Table \ref{GenericNorms}.  Those with the same color vanish at the same value of $\Delta$, where the states go null and shortening occurs.  The uncolored factors underneath are values where primaries in Table \ref{sconformalprimary} become singular.  At these values the extended modules discussed in Section \ref{extendedmodulesection} occur.}
\label{shortsmultiplets}
\end{center}
\end{figure}

Shortenings as they occur in the $r,\Delta$ plane are visualized in Figure \ref{shortsplanes}, and as they occur in the various levels of the multiplet in \ref{shortsmultiplets}.

\newpage
\begin{center}
\begin{longtable}{ |m{5cm}| m{9cm}|}
\caption{Norms of the conformal primaries in the $\{s\}_{\Delta, r}$ superconformal multiplet, up to an overall positive constant, as they appear in Table \ref{sconformalprimary}.  We have omitted totally symmetric products of $\delta$-functions in spinor indices for notational brevity.  
}
\label{GenericNorms}\\
\hline
\multicolumn{2}{|c|}{\textbf{Conformal Primary Norms: $\{s\}_{\Delta,r}$}}\\
\hline
{\textbf{Conformal Primary}} & {\textbf{Norm}}\\
\hline
\multicolumn{2}{|c|}{\textbf{Level 1}}\\
\hline
$[s - \frac{1}{2}]_{\Delta + \frac{1}{2}, r - 1}$ & $(\Delta +  r - s - 1)$\\
\hline
$[s - \frac{1}{2}]_{\Delta + \frac{1}{2}, r + 1}$ & $(\Delta- r - s - 1)$\\
\hline
$[s + \frac{1}{2}]_{\Delta + \frac{1}{2}, r - 1}$ & $(\Delta +  r + s)$\\
\hline
$[s + \frac{1}{2}]_{\Delta + \frac{1}{2}, r + 1}$ & $(\Delta - r + s)$\\
\hline
\multicolumn{2}{|c|}{\textbf{Level 2}}\\
\hline
$[s - 1]_{\Delta + 1, r}$ & $\frac{(\Delta + r - s - 1)(\Delta - r - s - 1)(\Delta - s)}{(\Delta - s - 1)}$\\
\hline
\multicolumn{2}{|c|}{$[s]^{(1,2)}_{\Delta + 1, r}$\,\,\,\, Graham Matrix  }\\
\hline
\multicolumn{2}{|c|}{$ \left(\begin{array}{cc} 8\Big(\Delta^2 - r^2 - s(s + 1)\frac{\Delta}{\Delta - 1}\Big) & -4\frac{(\Delta^2 + s\Delta - r\Delta + r)(\Delta - s + r - 1)}{(\Delta - 1)} \\-4\frac{(\Delta^2 + s\Delta - r\Delta + r)(\Delta + r - s - 1)}{(\Delta - 1)}&  2\frac{(\Delta  + r - s- 1)(\Delta(\Delta - r - 1) + s^2(2\Delta - 1) + s(2\Delta^2 - 2r\Delta + r - 1))}{s(\Delta - 1)} \end{array}\right)$}\\
\hline
$[s]^{(1,2)}_{\Delta + 1, r}$\,\,\,\, Graham Det. & $\frac{\Delta(\Delta + r - s - 1)(\Delta - r - s - 1)(\Delta  + r+ s)(\Delta  - r+ s)}{\Delta - 1}$\\
\hline
$[s]_{\Delta + 1, r - 2}$& $(\Delta + r - s - 1)(\Delta + r + s)$ \\
\hline
$[s]_{\Delta + 1, r + 2}$ & $(\Delta - r - s - 1)(\Delta - r + s)$ \\
\hline
$[s + 1]_{\Delta + 1, r}$ & $\frac{(\Delta + s + 1)(\Delta  + r+ s)(\Delta  - r+ s)}{\Delta + s}$ \\
\hline
\multicolumn{2}{|c|}{\textbf{Level 3}}\\
\hline
$[s - \frac{1}{2}]_{\Delta + \frac{3}{2}, r - 1}$ & $\frac{\Delta(\Delta - s)(\Delta + r - s - 1)(\Delta - r - s - 1)(\Delta  +  r+ s)}{(\Delta - \frac{1}{2})(\Delta - s - 1)}$\\
\hline
$[s - \frac{1}{2}]_{\Delta + \frac{3}{2}, r + 1}$ & $\frac{\Delta(\Delta - s)(\Delta + r - s - 1)(\Delta - r - s - 1)(\Delta  - r+ s)}{(\Delta - \frac{1}{2})(\Delta - s - 1)}$\\
\hline
$[s + \frac{1}{2}]_{\Delta + \frac{3}{2}, r - 1}$ & $\frac{\Delta(\Delta + s + 1)(\Delta +  r - s - 1)(\Delta + r + s)(\Delta - r + s)}{(\Delta - \frac{1}{2})(\Delta + s)}$\\
\hline
$[s + \frac{1}{2}]_{\Delta + \frac{3}{2}, r + 1}$ & $\frac{\Delta(\Delta + s + 1)(\Delta - r - s - 1)(\Delta + r + s)(\Delta - r + s)}{(\Delta - \frac{1}{2})(\Delta + s)}$\\
\hline
\multicolumn{2}{|c|}{\textbf{Level 4}}\\
\hline
$[s]_{\Delta + 2, r}$ & $\frac{(\Delta + \frac{1}{2})(\Delta - s)(\Delta + s + 1)(\Delta + r - s - 1)(\Delta - r - s - 1)(\Delta + r + s)(\Delta - r + s)}{(\Delta - s - 1)(\Delta + s)(\Delta - \frac{1}{2})}$\\
\hline
\end{longtable}
\end{center}

\clearpage

The only unitary short multiplets are the short massive multiplets that occur along the lines $\Delta=s+1\pm r$ at the boundary of the green unitary region in Figure \ref{shortsplanes}. They describe short multiplets for a massive spin $S=s+1$ particle whose structure is shown here. In this picture and those like it below, black states have positive norm, red states have negative norm, and green states are zero norm null states that decouple from the multiplet.   For the degenerate state at level 2, the colors correspond to the two eigenvalues of the Graham matrix. The placement of the states is in accord with Figure \ref{s1multipletgeneric}:
\be
\begin{array}{l} \left\{ S-1 \right\}_{r+S,r},\ |r|>0,\ S\geq 1 \\ {\rm spin\ }  S  {\rm\ short\ massive\ multiplet:} \end{array}\ \ \ \ \ \ \ \ \raisebox{-96pt}{\epsfig{file=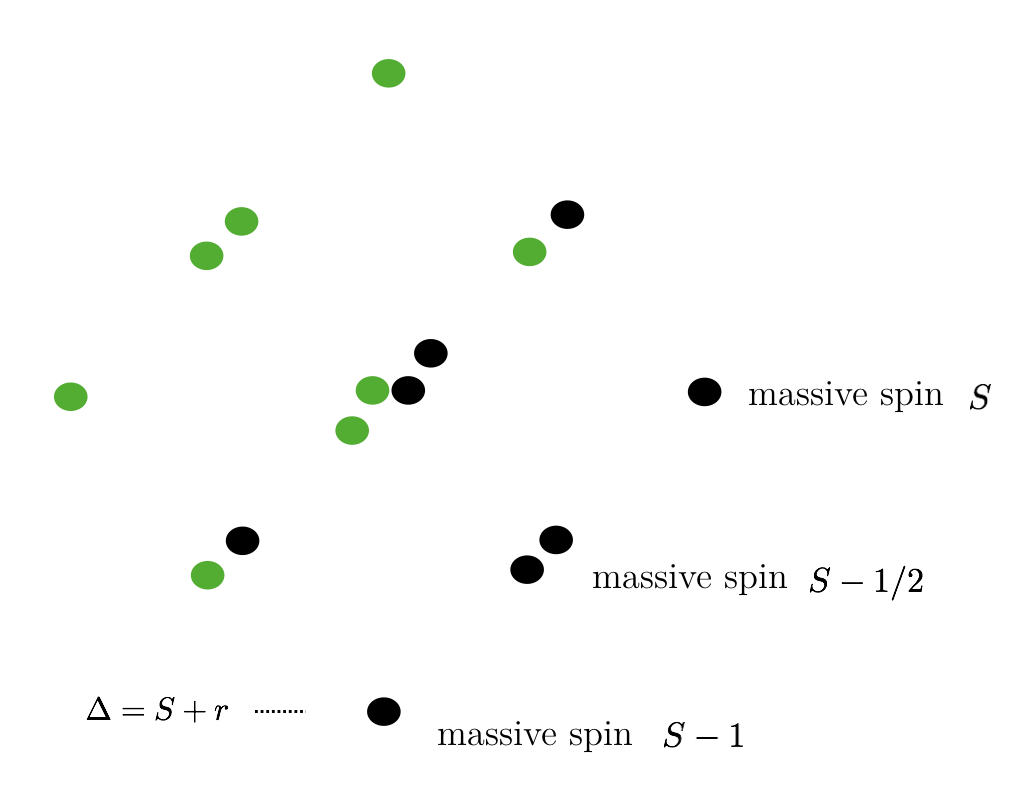,height=2.7in}}\label{massivemultipletpicse2}
\ee
This contains half as many degrees of freedom as the generic long massive multiplet.  As discussed in Section \ref{branchingsection}, this multiplet appears in the branching rules of a massive spin $S+1/2$ multiplet as it approaches its massless value.
When $r=0$, the intersection of the two lines $\Delta=s+1\pm r$ at the apex of the triangular green unitary region in Figure \ref{shortsplanes}, there is a further degeneration and we have the massless spin $S$ multiplet, shown in \eqref{PMmultipletpicse3}.
In the non-unitary region there are novel shortening conditions.  For example, along the $\Delta=-{1\over 2}$ shortening line of Figure \ref{shortsplanes}, we get a short multiplet where only the top state at level 4 decouples,
\be
\begin{array}{l} \left\{ s \right\}_{-{1\over 2},r}, \ \ s\geq 1,\ \ |r|<s-{1\over 2} , \end{array}\ \ \ \ \ \ \ \ \raisebox{-96pt}{\epsfig{file=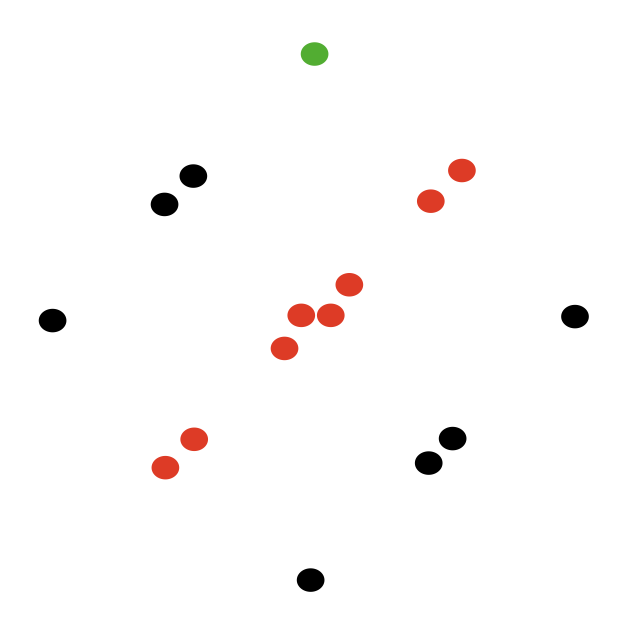,height=2.4in}}\label{massivemultipletpicse2}
\ee

\bigskip

\section{Extended modules\label{extendedmodulesection}}

In the various expressions for the conformal primaries in Section \ref{susconforepss}, the $P$ correction terms become singular for some values of $\Delta$.  At these values of $\Delta$, we have the phenomena of extended modules, which also occur in the ${\cal N}=1$ case  \cite{Garcia-Saenz:2018wnw} and in non-supersymmetric non-unitary CFT's \cite{Brust:2016gjy}.

The Hilbert space of a unitary CFT is spanned by primary operators and their descendants.   Often it is the case that the spectrum contains states which have zero norm which are both primary and descendant. For a unitary CFT, such states and their descendants are always orthogonal to every other state.  That is, they host their own null module, which may be consistently factored out, in which case we say that the multiplet shortens. 
Once such null states are factored out, the Hilbert space may be written as the direct sum
\begin{equation}
    \text{Ker}\, K \bigoplus_{n} \text{Im} \,P^{}\,.
    \label{directsum}
\end{equation}
Moreover, each factor may be further graded by the quantum numbers $\Delta, s, r$.  

For non-unitary CFT's, the story is more complicated.  For particular values of $\Delta$ a state may become both primary and descendant and thus have zero norm, however unlike in unitary theories, such states are generally {not orthogonal to every other state in the theory}, and thus do not host their own null module, in which case they cannot be factored out.  This is the hallmark of an extended module \cite{Brust:2016gjy}, and typically occurs when a conformal primary and a descendant (of a different primary) which are linearly independent for generic values of $\Delta$ degenerate into a single state for a particular value of $\Delta$.  In this case there exist other states which are inaccessible by taking linear combinations of primaries and descendants.  These inaccessible states must be neither primary nor descendant, because for these particular values of $\Delta, s$ the spaces $\text{Ker}\,K$ and $\text{Im}\,P$ coincide, and are spanned by the degenerate state.  This signifies a breakdown of the decomposition in~\eqref{directsum}.  When this happens, another zero-norm (non-null) basis state arises in the theory for these particular quantum numbers, spanning the missing direction (though not orthogonal to the degenerate states).  This state hosts its own module, which we call an extended module, which may be filled out by acting with $P$'s, and which cannot be factored out.

Below we will work out some examples of specific cases where the extended modules occur.
 As we will state later on, these extended modules appear in some of the superconformal multiplets which contain partially massless fields.  There are partially massless fields of spin $S$ with depth $t = S - 2$ which exist in superconformal multiplets whose primary has $\Delta = s$, $s=S-1$.  The simplest of these, in the sense that there are no higher-spin fields in the multiplet, occurs for $\Delta = s = 1$, which contains a partially massless spin-2 field with depth $t = 0$.  This superconformal multiplet contains an extended module which branches off at level 2 via the $\left[ 1\right]_{2,r}$  states.

\subsection{Extended module for $\{0\}_{0, r}$\label{extendeds0section}}

In the $\{0\}_{\Delta, r}$ superconformal multiplet, consider the space of states at level 2 with dimension $\Delta + 1$, $R$-charge $r$, and spin $1$.  This space is two dimensional and is spanned by the states
 \begin{align}
    &\ket{1}^{ab} \equiv \bar{Q}^{(a}Q^{b)}\ket{\Delta, r}  , \ \ \ \ket{2}^{ab}_{} \equiv P^{ab}\ket{\Delta, r}. \label{barebasiss0eqe}
\end{align}

For $\Delta\neq 0$, we can find an orthogonal basis consisting of the conformal primary at level 2 and the first descendant of the conformal primary at level 0,
\begin{align}
    &\ket{P}^{ab} \equiv \bar{Q}^{(a}Q^{b)}\ket{\Delta, r}  - \Big(\frac{\Delta - r}{\Delta}\Big)P^{a b}\ket{\Delta, r} ,\ \ \ \ket{D}^{ab}_{} \equiv P^{ab}\ket{\Delta, r}.
\end{align}

When $\Delta=0$ on the other hand, there is a degeneration: for $r=0$, both states \eqref{barebasiss0eqe} are primary, while for $r\not=0$ only $\ket{2}^{ab}$ is primary and is also descendent.  The decomposition \eqref{directsum} fails.  The Graham matrix of the basis \eqref{barebasiss0eqe} for $\Delta=0$ reads
\be  \left(\begin{array}{cc} {}_{cd} \la 1|1 \ra^{ab} & {}_{cd} \la 1|2 \ra^{ab} \\   {}_{cd} \la 2|1 \ra^{ab} & {}_{cd} \la 2|2 \ra^{ab} \end{array}\right)=\left(
\begin{array}{cc}
 -4 r(r+2)  & -4 r \\
 -4 r & 0 \\
\end{array}
\right)\delta\indices{^{(a} _c}\delta\indices{^{b)} _d} \,.  \label{grahammatrixseqe}
\ee

When $r=0$, the Graham matrix vanishes so both states are null and decouple. This is the vacuum multiplet.  This phenomena of extended states occurring within the null states often happens even in the unitary representations.  For example in the hypermultiplet $\{0\}_{1,1}$ and in the massless gauge multiplet $\{0\}_{0,1}$ there is an extended module occurring at level 2 where everything is null.

For $r\not=0$, the eigenvalues of \eqref{grahammatrixseqe} are both non-zero and of opposite sign to each other, so in diagonalizing \eqref{grahammatrixseqe} we would find one positive norm state and one negative norm state.  
Thus the Graham matrix is Lorentzian, and we can find a ``light-like'' basis consisting of two zero-norm states whose overlap with each other is non-zero:
\begin{align}
    &\ket{N_1}^{ab} \equiv \bar{Q}^{(a}Q^{b)}\ket{0, r}  - {r+2\over 2}P^{a b}\ket{0, r} ,\ \ \ \ket{N_2}^{ab}_{} \equiv P^{ab}\ket{0, r}.
\end{align}
\be  \left(\begin{array}{cc} {}_{cd} \la N_1|N_1 \ra^{ab} & {}_{cd} \la N_1|N_2 \ra^{ab} \\   {}_{cd} \la N_2|N_1 \ra^{ab} & {}_{cd} \la N_2|N_2 \ra^{ab} \end{array}\right)=\left(
\begin{array}{cc}
 0  & -4 r \\
 -4 r & 0 \\
\end{array}
\right)\delta\indices{^{(a} _c}\delta\indices{^{b)} _d} \,.  \label{grahammatrixseqe2}
\ee
$\ket{N_2}^{ab}_{}$ spans the subspace of primary and descendent, whereas $\ket{N_1}^{ab}$, termed an ``extension state'' in  \cite{Brust:2016gjy}, is neither primary nor descendant, and takes the place of the would-be primary at this level.  Though both states are zero-norm, they are not null because their overlap is non-zero and so there is no decoupling.

Note that from the point of view of the conformal algebra, a scalar primary with $\Delta=0$ is just the vacuum module with a single state because all the conformal descendants are null, leaving a short module with only a single state.  Here in the superconformal case with $r\not=0$, because of non-trivial mixing with $\ket{1}^{ab}$ and its descendants, the descendants, while still zero-norm, are no longer null and do not decouple, and so the module is no longer shortened.

One can also ask what happens if one acts on $\ket{N_1}^{ab}$ with $Q$.  For $\Delta = 0$, none of the basis vectors at level three degenerate, so the action of $Q$ simply moves one back into the original module.  There can however be instances for different combinations of $\Delta$ and $s$ where level 2 basis vectors degenerate, along with the basis vectors at higher levels. In such cases, the extended module spans multiple levels.

 \subsection{Extended module for $\{1\}_{1, r}$\label{extendeds1section}}
 As another example we look at the extended module that occurs in the $\{1\}_{1, r}$ representation.  For $r=0$ this is the extended module that appears in the PM spin-2 multiplet discussed in Section \ref{PMmultiplsecne}.  In this superconformal multiplet, consider the space of states at level 2 with dimension $ 2$, $R$-charge $r$, and spin $ 1$.  This space is three dimensional and is spanned by the states 
  \begin{align}
     &\ket{1}^{ab} \equiv \bar{Q}^{c}Q_{c}\ket{\Delta, r}^{ab}  , \ \ \ \ket{2}^{ab}_{} \equiv \bar{Q}_cQ^{(a}\ket{\Delta, r}^{b)c}   , \ \ \ \ket{3}^{ab} \equiv P\indices{^{(a}_c}\ket{\Delta, r}^{b)c}\, . \label{barebasiss1}
 \end{align}
 For $\Delta \neq 1$, there is a basis consisting of two conformal primaries at level 2 (the states $[s]^{(1)}_{\Delta + 1, r}$ and $[s]^{(2)}_{\Delta + 1, r}$ in Table \ref{sconformalprimary}) and the first spin-1 descendant of the conformal primary at level 0,
 \begin{align}
     &\ket{P_1}^{ab} \equiv \bar{Q}^{c}Q_{c}\ket{\Delta, r}^{ab} + \Big(\frac{2}{\Delta - 1}\Big)P\indices{^{(a}_c}\ket{\Delta, r}^{b)c}\, ,\nonumber \\
     &\ket{P_2}^{ab} \equiv \bar{Q}_cQ^{(a}\ket{\Delta, r}^{b)c} - \Big(\frac{\Delta - r}{\Delta - 1}\Big)P\indices{^{(a}_c}\ket{\Delta, r}^{b)c}\, ,\nonumber\\
     &\ket{D}^{ab} \equiv P\indices{^{(a}_c}\ket{\Delta, r}^{b)c}\, .
 \end{align}
 On the other hand, for $\Delta = 1$, there is degeneration, and we cannot construct such a basis. For $r\neq\pm2, \pm1$, $\ket{3}^{ab}$ is a primary, and it is also a descendant.  The Graham matrix for the basis \eqref{barebasiss1} for $\Delta = 1$ reads
 \begin{align}
     \left(\begin{array}{ccc} {}_{cd} \la 1|1 \ra^{ab} & {}_{cd} \la 1|2 \ra^{ab} & {}_{cd} \la 1|3 \ra^{ab} \\   {}_{cd} \la 2|1 \ra^{ab} & {}_{cd} \la 2|2 \ra^{ab} &{}_{cd} \la 2|3 \ra^{ab}\\   {}_{cd} \la 3|1 \ra^{ab} & {}_{cd} \la 3|2 \ra^{ab} &{}_{cd} \la 3|3 \ra^{ab} \end{array}\right) = \left(\begin{array}{ccc} -8(r^2 + 1) & 4(r^2 - 4 r - 1) & -8 \\ 4(r^2 - 4 r - 1) &-2(r - 1)(3r - 1) &-4(r - 1)\\-8 & -4(r - 1) & 0 \end{array}\right)\delta\indices{^{(a} _c}\delta\indices{^{b)} _d} \,.
     \label{GrahamBareBasiss1}
 \end{align}

 The determinant of this matrix is $128(r+1)(r-1)(r+2)(r-2)$, so for the special cases of $r = \pm1, \pm2 $, this Graham matrix has a vanishing eigenvalue.  The remaining two eigenvalues are nonvanishing and of opposite sign.  For these particular values of $r$, we have an extended module and a null module.  Looking at Figure \ref{shortsplanes}, we see that $r=\pm 1$ are the cases where the line $\Delta=1$ intersects the $\Delta=s+1\mp r$ shortening line, and $r=\pm 2$ are the cases where the line $\Delta=1$ intersects the $\Delta=\pm r-sr$ shortening line, so we are seeing these shortenings impact the extended module.  We will return to these cases at the end of this section.  

 On the other hand, for generic values of $r$ (in particular, this applies to the PM spin-2 multiplet discussed in Section \ref{PMmultiplsecne} which has $r=0$), the Graham matrix always has two positive and one negative eigenvalue, or two negative and one positive eigenvalue.  We can find a basis of zero-norm states whose overlap with each other is non-zero:   
 \begin{align}
     &\ket{N_1}^{ab} = \bar{Q}^{c}Q_{c}\ket{\Delta, r}^{ab} - \frac{1 + r^2}{2}P\indices{^{(a}_c}\ket{\Delta, r}^{b)c}\, , \nonumber \\
     &\ket{N_2}^{ab} = \bar{Q}_{c}Q^{(a}\ket{\Delta, r}^{b)c} + \frac{1 - 3r}{4}P\indices{^{(a}_c}\ket{\Delta, r}^{b)c}\, ,\nonumber\\
     &\ket{N_3}^{ab} = P\indices{^{(a}_c}\ket{\Delta, r}^{b)c}\, ,
 \end{align}
 \begin{align}
     &\left(\begin{array}{ccc} {}_{cd} \la N_1|N_1 \ra^{ab} & {}_{cd} \la N_1|N_2 \ra^{ab} & {}_{cd} \la N_1|N_3 \ra^{ab} \\   {}_{cd} \la N_2|N_1 \ra^{ab} & {}_{cd} \la N_2|N_2 \ra^{ab} &{}_{cd} \la N_2|N_3 \ra^{ab}\\   {}_{cd} \la N_3|N_1 \ra^{ab} & {}_{cd} \la N_3|N_2 \ra^{ab} &{}_{cd} \la N_3|N_3 \ra^{ab} \end{array}\right) \nonumber \\
     &\hspace{2 cm}= \left(\begin{array}{ccc} 0 & 2(r + 1)(r + 2)(r - 2) & -8 \\ 2(r + 1)(r + 2)(r - 2) &0 &-4(r - 1)\\-8 & -4(r - 1) & 0 \end{array}\right)\delta\indices{^{(a} _c}\delta\indices{^{b)} _d}. 
     \label{zeronormbasis}
 \end{align}
 $\ket{N_3}^{ab}$ spans the subspace of primaries and descendants.  On the other hand, $\ket{N_1}^{ab}$ and $\ket{N_2}^{ab}$ are extension states, and are neither primary nor descendant.  They take the place of the would-be pair of primaries at this level, and host the extended module, which can be filled out by the action of $P$'s on linear combinations of these states.

 Now we will return to the cases $r = \pm1, \pm2 $.  In each of these cases, the Graham matrix \eqref{GrahamBareBasiss1} has a zero eigenvalue, and two additional eigenvalues of opposite sign.  This indicates that in each case there is a basis of zero-norm vectors, one of which will have zero overlap with the others.  The other two will have non-zero overlap with each other. The basis vector with zero-overlap is a true null state, and may be consistently factored out.  On the other hand, one of the remaining two will be an extension state, and host an extended module.  The remaining state will span the space of states which are simultaneously primary and descendant.  

 \textbf{Cases $r = \pm2, -1$:} We can handle these cases together because for each of these values of $r$, the $(1, 2), (2, 1)$ entries of the Graham matrix \eqref{zeronormbasis} vanish.  Thus, any linear combination of $\ket{N_1}$ and $\ket{N_2}$ has zero norm by default, and we need only find a linear combination which is both a conformal primary and orthogonal to $\ket{N_3}$.  A particular choice of basis is
 \begin{align}
     &\ket{\mathcal{P}_1}^{ab} = \ket{N_1}^{ab} + \frac{2}{1 - r}\ket{N_2}^{ab}\,  \nonumber \\
     &= \bar{Q}^{c}Q_{c}\ket{\Delta, r}^{ab} + \frac{2}{1 - r}\bar{Q}_{c}Q^{(a}\ket{\Delta, r}^{b)c} - \frac{r(r + 1)(r - 2)}{2(r - 1)}P\indices{^{(a}_c}\ket{\Delta, r}^{b)c}\, ,\nonumber \\
     &\ket{N_2}^{ab} = \bar{Q}_{c}Q^{(a}\ket{\Delta, r}^{b)c} + \frac{1 - 3r}{4}P\indices{^{(a}_c}\ket{\Delta, r}^{b)c}\, ,\nonumber\\
     &\ket{N_3}^{ab} = P\indices{^{(a}_c}\ket{\Delta, r}^{b)c}\, ,
 \end{align}
 \begin{align}
     \left(\begin{array}{ccc} {}_{cd} \la \mathcal{P}_1|\mathcal{P}_1 \ra^{ab} & {}_{cd} \la \mathcal{P}_1|N_2 \ra^{ab} & {}_{cd} \la \mathcal{P}_1|N_3 \ra^{ab} \\   {}_{cd} \la N_2|\mathcal{P}_1 \ra^{ab} & {}_{cd} \la N_2|N_2 \ra^{ab} &{}_{cd} \la N_2|N_3 \ra^{ab}\\   {}_{cd} \la N_3|\mathcal{P}_1 \ra^{ab} & {}_{cd} \la N_3|N_2 \ra^{ab} &{}_{cd} \la N_3|N_3 \ra^{ab} \end{array}\right)= \left(\begin{array}{ccc} 0 & 0 & 0 \\ 0 &0 &-4(r - 1)\\0 & -4(r - 1) & 0 \end{array}\right)\delta\indices{^{(a} _c}\delta\indices{^{b)} _d},\ \ r = \pm2, -1. 
 \end{align}
 The state $\ket{\mathcal{P}_1}$ is null and a conformal primary, and hosts its own null module which may be factored out.  On the other hand, the extension state $\ket{N_2}$ is zero-norm but not null, and hosts an extension module, taking the place of the second conformal primary. Finally, $\ket{N_3}$ spans the space of states which are simultaneously primary and descendant.   

 \textbf{Case $r = 1$:} In this case, the $(2, 3)$, $(3, 2)$ entries of the Graham matrix \eqref{zeronormbasis} vanish, which suggests that we should construct the conformal primary which hosts the null module by taking a linear combination of $\ket{N_2}$ and $\ket{N_3}$.  Moreover, for $r = 1$ both $\ket{N_2}$ and $\ket{N_3}$ are primary, so we need only worry about constructing a state which is null. A particular choice of basis is
 \begin{align}
     &\ket{N_1}^{ab} = \bar{Q}^{c}Q_{c}\ket{\Delta, 1}^{ab} -P\indices{^{(a}_c}\ket{\Delta, 1}^{b)c}\, ,\nonumber \\
     &\ket{\mathcal{P}_2}^{ab} = \ket{N_2}^{ab} - \frac{3}{2}\ket{N_3}^{ab} = \bar{Q}_{c}Q^{(a}\ket{\Delta, 1}^{b)c} - 2 P\indices{^{(a}_c}\ket{\Delta, 1}^{b)c}\, ,\nonumber \\
     &\ket{N_3}^{ab} = P\indices{^{(a}_c}\ket{\Delta, 1}^{b)c}\, ,
 \end{align}
 \begin{align}
     \left(\begin{array}{ccc} {}_{cd} \la N_1|N_1 \ra^{ab} & {}_{cd} \la N_1|\mathcal{P}_2 \ra^{ab} & {}_{cd} \la N_1|N_3 \ra^{ab} \\   {}_{cd} \la \mathcal{P}_2|N_2 \ra^{ab} & {}_{cd} \la \mathcal{P}_2|\mathcal{P}_2 \ra^{ab} &{}_{cd} \la \mathcal{P}_2|N_3 \ra^{ab}\\   {}_{cd} \la N_3|N_1 \ra^{ab} & {}_{cd} \la N_3|\mathcal{P}_2 \ra^{ab} &{}_{cd} \la N_3|N_3 \ra^{ab} \end{array}\right)= \left(\begin{array}{ccc} 0 & 0 & -8 \\ 0 &0 &0\\-8 & 0 & 0 \end{array}\right)\delta\indices{^{(a} _c}\delta\indices{^{b)} _d}. 
 \end{align}
 Similar to the prior case, the state $\ket{\mathcal{P}_2}$ is null and a conformal primary, a hosts its own null module which may be factored out.  The extension state is $\ket{N_1}$, and $\ket{N_3}$ spans the space of states which are simultaneously primary and descendant.

 Finally, note that at level 3 in this superconformal multiplet, there is no degeneration for $\Delta = s = 1$. Therefore, acting with $Q$ on the extension states and their $P$-descendants moves one back into the original module.  This applies for all values of $r$.

\bigskip

\section{Partially massless multiplets\label{PMmultiplsecne}}

In the boundary CFT, partially massless particles of spin $S$ and depth $t$ correspond to short multiplets with conformal dimension $\Delta=t+2 $, which have a null descendent at level $S-t$.

The simplest partially massless SUSY multiplets are those whose highest spin component is a PM field.  Thus the superconformal primary should have spin $S-1$ and weight $\Delta=t+1$.  If we don't want the PM field to be charged under $R$-symmetry then we should take $r=0$.  Thus the ${\cal N}=2$ SUSY multiplet of interest for a PM field of spin $S$ depth $t$ is
\be \left\{ S-1\right\}_{t+1,0}.\ee
The PM field of interest occurs at level 2, with weight $\Delta+1=t+2$. If we compare this with the shortening conditions found above, we see that there are short supermultiplets where the highest-spin state in the multiplet is a partially massless state of depth $t=S-2$.

The various partially massless multiplets described in the paragraph above are as follows.  In all the pictures, black states have positive norm, red states have negative norm, and green states are zero norm null states that decouple from the multiplet.   For the degenerate state at level 2, the colors correspond to the two eigenvalues of the Graham matrix.  Blue states are where the extended modules described in Section \ref{extendedmodulesection} occur.  The placement of the states is in accord with Figures \ref{s0multipletgeneric}, \ref{s1halfmultipletgeneric}, \ref{s1multipletgeneric}.  The partially massless fields of spin $S$ and depth $t$ are labelled by the notation $(S,t)$.

\textbf{Spin 1 massless:}  The only partially massless point for spin 1 is the standard massless value at $t=0$, where we hit the unitary shortening condition at $\Delta=1$.  This multiplet is $\{ 0\}_{1,0}$, shown here: 
\be
\left\{ 0\right\}_{1,0}\ {\rm spin\ 1\ massless\ multiplet:}\ \ \ \ \ \ \ \ \raisebox{-96pt}{\epsfig{file=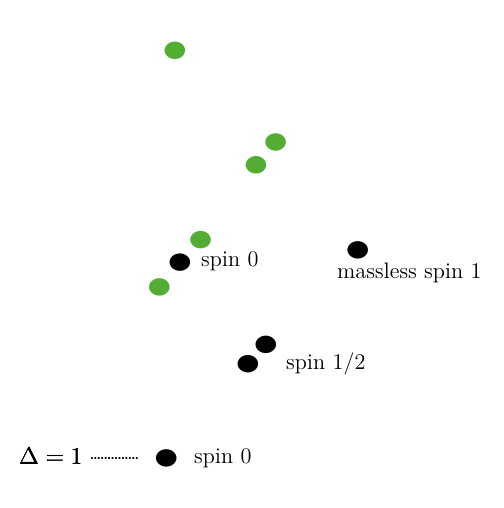,height=2.7in}}\label{PMmultipletpicse1}
\ee
The bulk has 4 propagating bosonic degrees of freedom and the same number of propagating fermionic degrees of freedom.

\textbf{Spin 3/2 massless:}  The only partially massless point for spin 3/2 is the standard massless value at $t=1/2$, where we hit the unitary shortening condition at $\Delta=3/2$. This multiplet is $\{ {1\over 2} \}_{{3\over 2},0}$, shown here: 
\be
\left\{ {1\over 2} \right\}_{{3\over 2},0}\ {\rm spin\ 3/2\ massless\ multiplet:}\ \ \ \ \ \ \ \ \raisebox{-96pt}{\epsfig{file=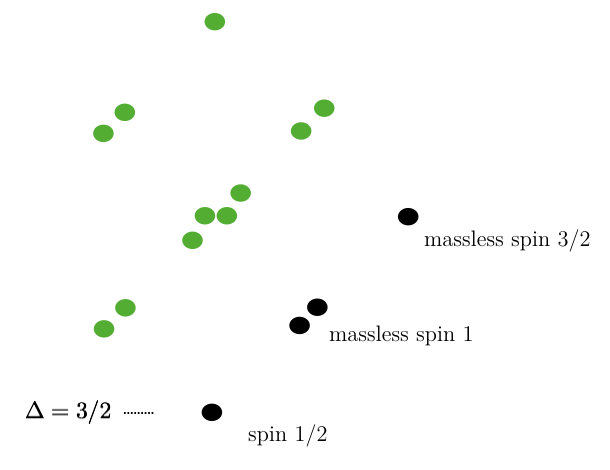,height=2.7in}}\label{PMmultipletpicse2}
\ee
The bulk has 4 propagating bosonic degrees of freedom and the same number of propagating fermionic degrees of freedom.

\textbf{Spin $S\geq 2$ massless:} For spin $S\geq 2$ the massless case is $t=S-1$ where we hit the unitary shortening condition at $\Delta=s+1$.  This multiplet is $\{ S-1 \}_{S,0}$, shown here: 
\be
\left\{ S-1 \right\}_{S,0}\ {\rm spin\ } S {\rm \ massless\ multiplet:}\ \ \ \ \ \ \ \ \raisebox{-96pt}{\epsfig{file=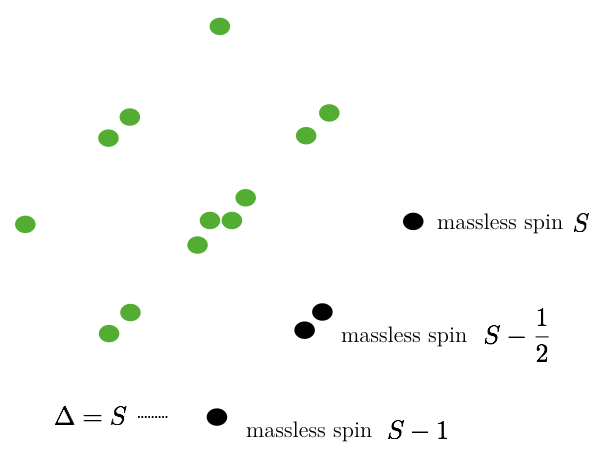,height=2.7in}}\label{PMmultipletpicse3}
\ee
It contains only massless fields.  The bulk has 4 propagating bosonic degrees of freedom and the same number of propagating fermionic degrees of freedom.
 
\textbf{Spin $2$, $t=0$ PM short multiplet:} Here we hit the non-unitary level 2 shortening condition given by $\Delta = s$ which also coincides with the extended module condition.  This multiplet is $\{ 1 \}_{{1},0}$, shown here: 
\be
\left\{ 1 \right\}_{1,0}\ {\rm spin\ } 2 {\rm \ PM:}\ \ \ \ \ \ \ \ \raisebox{-96pt}{\epsfig{file=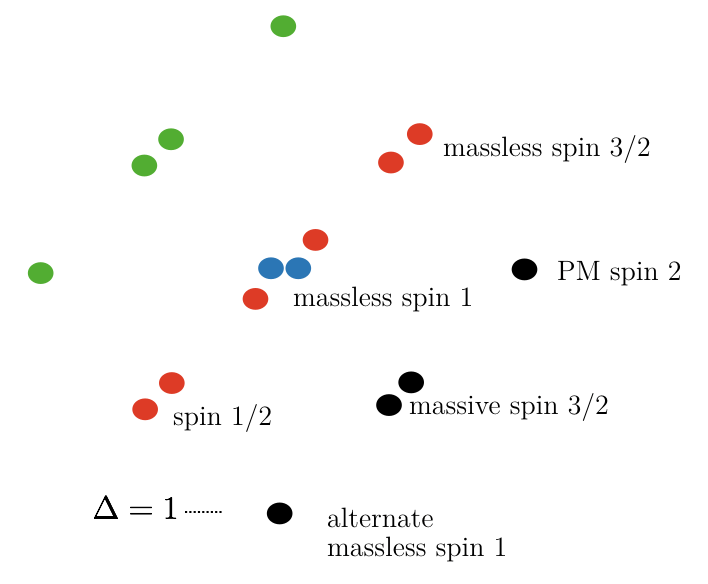,height=2.7in}}\label{PMmultipletpicse7}
\ee
   This short multiplet  includes massless and massive fields in addition to the partially massless spin-2.  It also contains the field $\left[ 1\right]_{0}$, obtained by performing an alternate quantization on a bulk massless spin-1.  Some of the other massless spin-1's live in an extended module.  If we demand that the multiplet has an equal number of bosonic and fermionic degrees of freedom, then the alternately quantized photon must carry 4 propagating degrees of freedom.  Given this,
the bulk has $16$ propagating bosonic degrees of freedom and the same number of propagating fermionic degrees of freedom.

 \textbf{Spin $5/2$, $t=1/2$ PM short multiplet:}  Here we hit the non-unitary level 2 shortening condition given by $\Delta = s$.  This multiplet is $\{ {3\over 2} \}_{{3\over 2},0}$, shown here: 
\be
\left\{{3\over 2} \right\}_{{3\over 2},0}\ {\rm spin\ } {5\over 2},\ t={1\over 2}{\rm \ PM:}\ \ \ \ \ \ \ \ \raisebox{-96pt}{\epsfig{file=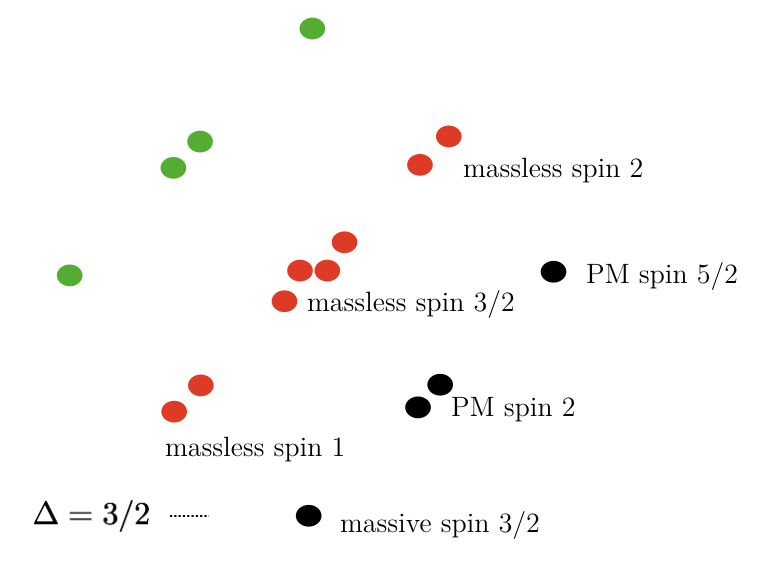,height=2.7in}}\label{PMmultipletpicse9}
\ee
   It contains PM fields and massless fields, and a single massive field of spin 3/2.   The bulk has $16$ propagating bosonic degrees of freedom and the same number of propagating fermionic degrees of freedom.

\textbf{Spin $S\geq 3$, $t=S-2$ PM short multiplet:}  At the next PM point beyond the massless point we hit the non-unitary level 2 shortening condition given by $\Delta = s$. This multiplet is $\{ S-1 \}_{S-1,0}$, shown here: 
\be
\left\{ S-1 \right\}_{S-1,0}\ {\rm Spin\ } S\geq 3,\ t=S-2 {\rm\ PM\ short }\ \ \ \ \ \ \ \ \raisebox{-96pt}{\epsfig{file=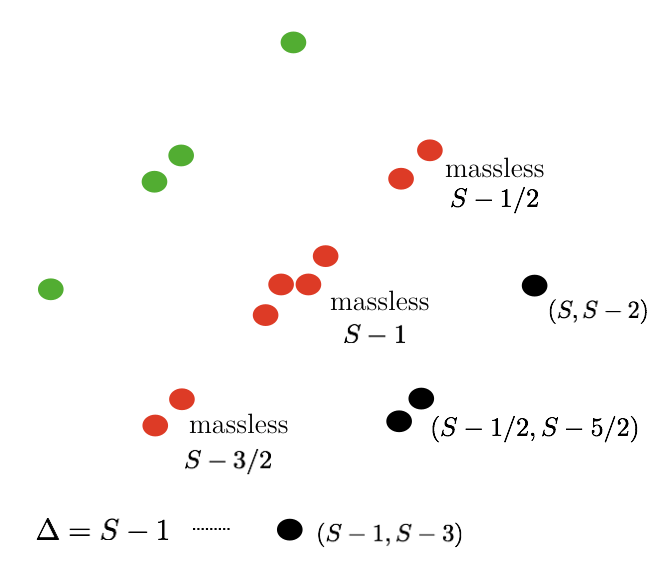,height=2.7in}}\label{PMmultipletpicse4}
\ee
  It contains massless and  depth $t=S-2$ partially massless fields.  The bulk has 16 propagating bosonic degrees of freedom and the same number of propagating fermionic degrees of freedom.

\textbf{Spin $S\geq 4$, $1\leq t \leq S-3$ PM long multiplet:} For these PM values we have no shortening condition and the PM multiplet is a generic long multiplet.  This multiplet is $\{ S-1 \}_{t+1,0}$, shown here: 
\be
\left\{ S-1 \right\}_{t+1,0}\ {\rm spin\ } S\geq 4,\ 1\leq t\leq S-3 {\rm \ PM\ long:}\ \ \ \ \ \ \ \ \raisebox{-96pt}{\epsfig{file=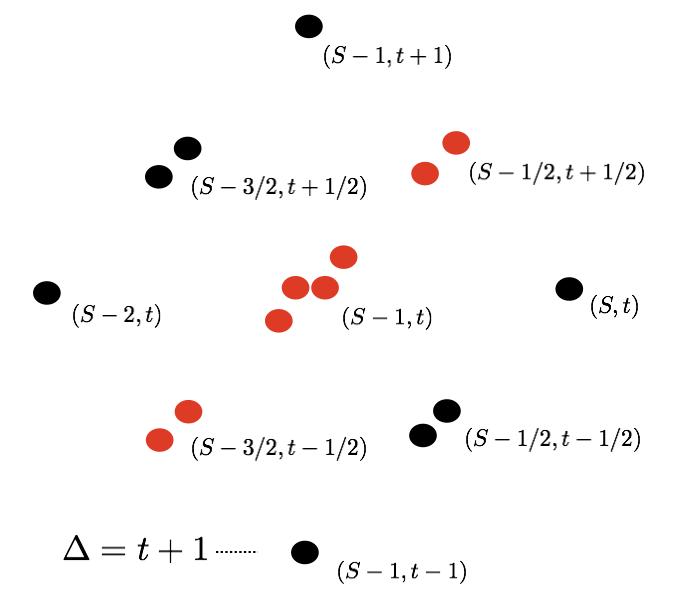,height=2.7in}}\label{PMmultipletpicse5}
\ee
   It contains only PM fields, and massless fields only if $t=S-3$.  The bulk has $16(S-t-1)$ propagating bosonic degrees of freedom and the same number of propagating fermionic degrees of freedom.

\textbf{Spin $S\geq 3$, $t=0$ PM multiplet:}  This is the minimal depth bosonic multiplet.   There is no shortening condition and it is a generic long multiplet.  This multiplet is $\{ S-1 \}_{{1},0}$, shown here: 
\be
\left\{ S-1 \right\}_{1,0}\ {\rm spin\ } S\geq3,\ t=0 {\rm \ PM:}\ \ \ \ \ \ \ \ \raisebox{-96pt}{\epsfig{file=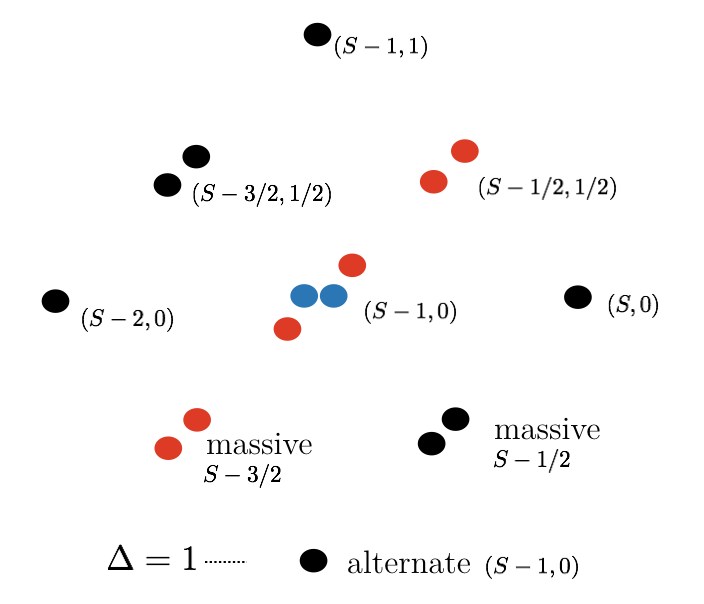,height=2.7in}}\label{PMmultipletpicse6}
\ee
   It contains PM fields (which are massless only for $S=3$) and massive fields.   In addition, it contains the field $\left[ S-1\right]_{0}$.  This is obtained by performing an alternate quantization on a bulk PM field of spin $S-1$ with $t=0$.  The bulk degree of freedom counting of this alternately quantized field is unclear, but if we demand that the multiplet has an equal number of bosonic and fermionic degrees of freedom, we can infer that it must propagate $2s$ degrees of freedom.  Given this,
the bulk has $16(S-1)$ propagating bosonic degrees of freedom and the same number of propagating fermionic degrees of freedom.

\newpage

\textbf{Spin $S\geq 7/2$, $t=1/2$ PM multiplet:}  This is the minimal depth fermionic multiplet.   There is no shortening condition and it is a generic long multiplet.  This multiplet is \linebreak $\{ S-1 \}_{{3\over 2},0}$, shown here: 
\be
\left\{ S-1 \right\}_{{3\over 2},0}\ {\rm spin\ } S\geq 7/2,\ t=1/2 {\rm \ PM:}\ \ \ \ \ \ \ \ \raisebox{-96pt}{\epsfig{file=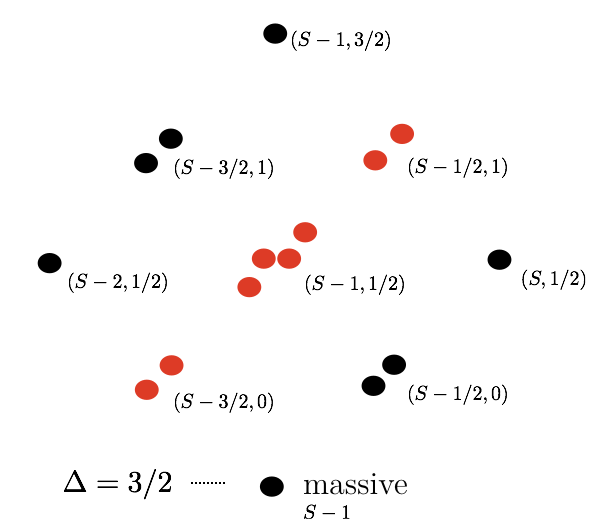,height=2.7in}}\label{PMmultipletpicse8}
\ee
   It contains PM fields and a single massive field .  The bulk has $16(S-t-1)$ propagating bosonic degrees of freedom and the same number of propagating fermionic degrees of freedom.

\begin{figure}[h!]
\begin{center}
\epsfig{file=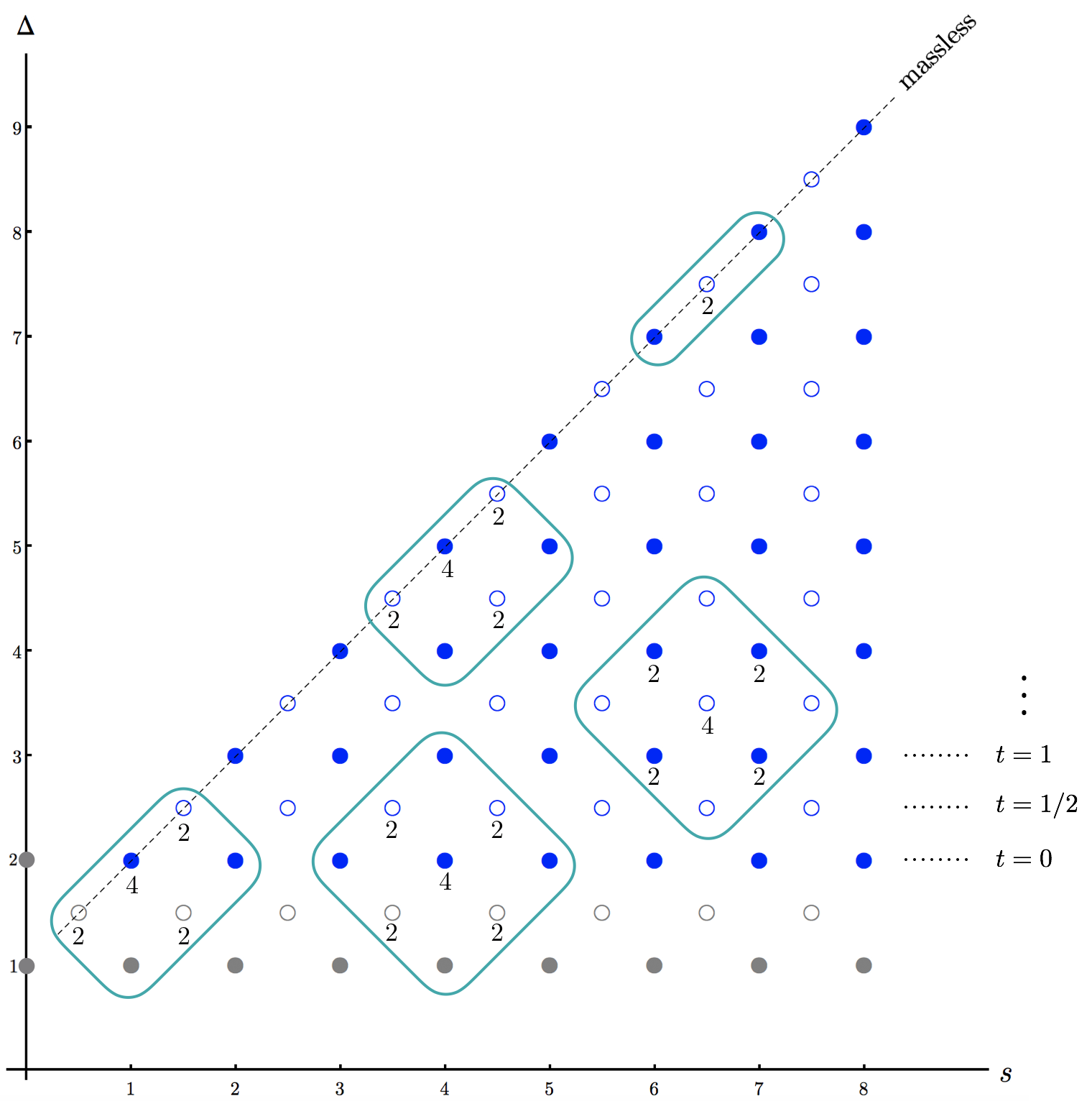,height=4.5in}
\caption{\small  ${\cal N}=2$ partially massless multiplets.   Filled circles are bosons, hollow circles are fermions.   The blue circles are all the partially massless points.  The grey circles, such as those on the bottom two rows, are massive states that participate in some of the multiplets.  The generic PM multiplets are $2\times 2$ diamonds, like the example showing the $\left\{6  {1\over 2} \right\}_{2{1\over 2},0 }$ spin $7  {1\over 2}$ depth $t= 3/2$ PM multiplet (the numbers inside the diamond show the multiplicities of the fields which occur more than once in the representation).  As the diamonds approach the massless line $\Delta=s+1$ at the top, we get the shortened PM multiplets, like the $\left\{4 \right\}_{4,0 }$ spin $5$ depth $t=3$ example shown, and the massless multiplets, like the $\left\{6 \right\}_{7,0 }$ spin $7$ example shown.  When the diamond reaches the bottom line $\Delta=1$, we get the PM multiplets containing extended modules, like the $\left\{4 \right\}_{1,0 }$ spin $5 $ depth $t= 0$ example shown.  These are the PM multiplets which cross the $\Delta=3/2$ divide between ordinary and alternate quantization of bulk AdS$_4$ fields.  The case where we get both a shortening and extended modules is the PM spin 2 multiplet $\left\{1 \right\}_{1,0 }$, also shown.}
\label{allPM2fig}
\end{center}
\end{figure}

The various PM fields and how they fit into the multiplets described in this section are visualized in Figure \ref{allPM2fig}.

\subsection{Branching rules\label{branchingsection}}

As a generic massive multiplet approaches the values of a partially massless or other multiplet containing null states, the null states decouple into their own multiplet and we have a branching rule.  Here we determine the branching rules for the PM multiplets described in Section \ref{PMmultiplsecne}.

In the partially massless case, the null states corresponding to the partially conserved operators can be thought of as the gauge modes of the corresponding bulk partially massless field. 
As a conformal primary approaches a partially massless value, we have the branching rule \cite{Bonifacio:2018zex}
\be \left[ s\right]_{\Delta}\underset{\Delta\rightarrow t+2}{\rightarrow}  \left[ s\right]_{t+2} \oplus \left[ t\right]_{s+2}, \ee
where the first summand is the PM field and the second summand is the gauge mode.  From this we can deduce the following branching rules of the PM multiplets:

\textbf{Spin 1 massless:}  The longitudinal modes form two short hypermultiplets of opposite $r$ charge,
\be \left\{  0 \right\}_{\Delta,0} \underset{\Delta\rightarrow 1}{\rightarrow}  \left\{  0 \right\}_{1,0}\oplus   \left\{0  \right\}_{2,2} \oplus   \left\{0  \right\}_{2,-2}  \ee

\textbf{Spin 3/2 massless:}  The longitudinal modes form two short massive spin 1 reps of opposite $r$ charge,
\be \left\{  {1\over 2} \right\}_{\Delta,0} \underset{\Delta\rightarrow {3\over 2}}{\rightarrow}  \left\{  {1\over 2} \right\}_{{3\over 2},0}\oplus   \left\{  0 \right\}_{2,1}\oplus   \left\{  0 \right\}_{2,-1} \ee

\textbf{Spin $S\geq 2$ massless:}  The longitudinal modes form two short massive reps of opposite $r$ charge,
\be \left\{  S-1 \right\}_{\Delta,0} \underset{\Delta\rightarrow S}{\rightarrow}  \left\{S-1 \right\}_{S,0}\oplus   \left\{  S-{3\over 2} \right\}_{S+{1\over 2},1}\oplus   \left\{  S-{3\over 2} \right\}_{S+{1\over 2},-1} \ee

\textbf{Spin $2$, $t=0$ PM short multiplet:}  The longitudinal modes form a long massive scalar multiplet, 
\be \left\{  1 \right\}_{\Delta,0} \underset{\Delta\rightarrow 1}{\rightarrow}  \left\{  1 \right\}_{1,0}\oplus   \left\{ 0 \right\}_{2,0} \ee
Note that the state $\left[1\right]_1$ does not spin off any gauge modes.

\textbf{Spin $5/2$, $t=1/2$ PM short multiplet:} The longitudinal mode is a long massive spin 3/2 multiplet,
\be \left\{   {3\over 2}\right\}_{\Delta,0} \underset{\Delta\rightarrow  {3\over 2}}{\rightarrow}  \left\{  {3\over 2} \right\}_{{3\over 2},0}\oplus   \left\{ {1\over 2}  \right\}_{{5\over 2},0} \ee

\textbf{Spin $S\geq 3$, $t=s-2$ PM short multiplet:}  The longitudinal modes form a generic long massive multiplet,
\be \left\{  S-1 \right\}_{\Delta,0} \underset{\Delta\rightarrow S-1}{\rightarrow}  \left\{  S-1 \right\}_{S-1,0}\oplus   \left\{ S-2  \right\}_{S,0} \ee

\textbf{Spin $S\geq 4$, $1\leq t \leq S-3$ PM long multiplet:}   The longitudinal modes form a generic long massive multiplet, all of whose members are PM gauge modes,
\be \left\{  S-1 \right\}_{\Delta,0} \underset{\Delta\rightarrow t+1}{\rightarrow}  \left\{  S-1 \right\}_{t+1,0}\oplus   \left\{ t  \right\}_{S,0} \ee

\textbf{Spin $S\geq 3$, $t=0$ PM multiplet:} The longitudinal modes form a generic long massive scalar multiplet, all of whose members are PM gauge modes,
\be \left\{  S-1 \right\}_{\Delta,0} \underset{\Delta\rightarrow 1}{\rightarrow}  \left\{  S-1 \right\}_{1,0}\oplus   \left\{ 0 \right\}_{S,0} \ee
Note that the state $\left[S-1\right]_1$ does not spin off any gauge modes, even though it develops a null descendent from the point of view of the conformal algebra.  This is because that null state is not null with respect to the full superconformal algebra, because it is part of the extended module that develops.

\textbf{Spin $S\geq 7/2$, $t=1/2$ PM multiplet:}  The longitudinal mode is a long massive spin 3/2 multiplet, all of whose members are PM gauge modes,
\be \left\{   S-1 \right\}_{\Delta,0} \underset{\Delta\rightarrow  {3\over 2}}{\rightarrow}  \left\{  S-1 \right\}_{{3\over 2},0}\oplus   \left\{ {1\over 2}  \right\}_{S,0} \ee

\bigskip

\subsection{Reduction to ${\cal N}=1$}

All of the ${\cal N}=2$ multiplets can be decomposed into the ${\cal N}=1$ multiplets classified in \cite{Garcia-Saenz:2018wnw} by simply finding the unique ${\cal N}=1$ multiplets which combine into the desired ${\cal N}=2$ multiplet. The generic massive spin $s\geq 1$ long multiplet splits as
\be \left\{s\right\}_{\Delta}^{{\cal N}=2}= \left\{s\right\}_{\Delta}^{{\cal N}=1}\oplus \left\{s+{1\over 2}\right\}_{\Delta+{1\over 2}}^{{\cal N}=1}\oplus\left\{s-{1\over 2}\right\}_{\Delta+{1\over 2}}^{{\cal N}=1}\oplus\left\{s\right\}_{\Delta+1}^{{\cal N}=1}. \ee

The short PM multiplets are also easily reduced:

\textbf{Spin 1 massless:}  The massless spin $1$ multiplet splits into two ${\cal N}=1$ multiplets, a massless spin $1$ multiplet and a massive scalar multiplet,
\be \left\{0 \right\}_{1,0}^{{\cal N}=2}= \left\{{1\over 2}\right\}_{{3\over 2}}^{{\cal N}=1}\oplus \left\{0 \right\}_{1}^{{\cal N}=1}    \ee

\textbf{Spin 3/2 massless:}  The massless spin $3/2$ multiplet splits into two massless ${\cal N}=1$ multiplets, a massless spin $3/2$ multiplet and a massless spin $1$ multiplet,
\be \left\{{1\over 2}\right\}_{{3\over 2},0}^{{\cal N}=2}= \left\{1\right\}_{2}^{{\cal N}=1} \oplus \left\{{1\over 2}\right\}_{{3\over 2}}^{{\cal N}=1} \ee

\textbf{Spin $S\geq 2$ massless:}  The massless spin $S$ multiplet splits into two massless ${\cal N}=1$ multiplets, a massless spin $S$ multiplet and a massless spin $S-{1\over 2}$ multiplet,
\be \left\{S-1\right\}_{S,0}^{{\cal N}=2}= \left\{S-{1\over 2}\right\}_{S+{1\over 2}}^{{\cal N}=1} \oplus \left\{S-1\right\}_{S}^{{\cal N}=1} \ee

\textbf{Spin $2$, $t=0$ PM short multiplet:}   The PM spin-2 multiplet splits as,
\be \left\{1\right\}_{1,0}^{{\cal N}=2}= \left\{1\right\}_{1}^{{\cal N}=1}\oplus \left\{{3\over 2}\right\}_{{3\over 2}}^{{\cal N}=1}\oplus\left\{{1\over 2}\right\}_{{3\over 2}}^{{\cal N}=1}\oplus\left\{{1 }\right\}_{{ 2}}^{{\cal N}=1}. \ee
The extended module and alternate quantized photon go into $ \left\{1\right\}_{1}^{{\cal N}=1}$.  Apart from this we have a PM spin-2 multiplet a massless spin 3/2 and a spin 1 multiplet on the ${\cal N}=1$ side.

\textbf{Spin $5/2$, $t=1/2$ PM short multiplet:} 

\be \left\{{3\over 2} \right\}_{{3\over 2},0}^{{\cal N}=2}= \left\{{3\over 2}\right\}_{{3\over 2}}^{{\cal N}=1}\oplus \left\{2\right\}_{2}^{{\cal N}=1}\oplus\left\{1\right\}_{2}^{{\cal N}=1}\oplus\left\{{3\over 2}\right\}_{{5\over 2}}^{{\cal N}=1}. \ee
PM spin-2 multiplet, PM spin 5/2 multiplet, massless spin 3/2 and massless spin 2.

\textbf{Spin $S\geq 3$, $t=S-2$ PM short multiplet:}  The short PM multiplets each split into two partially massless ${\cal N}=1$ multiplets and two massless ${\cal N}=1$ multiplets,
\be  \left\{  S-1 \right\}_{S-1,0}^{{\cal N}=2}= \left\{  S-1 \right\}_{S-1}^{{\cal N}=1} \oplus  \left\{  S-{1\over 2} \right\}_{S-{1\over 2}}^{{\cal N}=1}\oplus  \left\{  S-{3\over 2} \right\}_{S-{1\over 2}}^{{\cal N}=1}  \oplus\left\{  S-1 \right\}_{S}^{{\cal N}=1} \,.  \ee

\textbf{Spin $S\geq 4$, $1\leq t \leq S-3$ PM long multiplet:}   The long PM multiplets each split into four different partially massless ${\cal N}=1$ multiplets,
\be  \left\{  S-1 \right\}_{t+1,0}^{{\cal N}=2}= \left\{  S-1 \right\}_{t+1}^{{\cal N}=1} \oplus  \left\{  S-{1\over 2} \right\}_{t+{3\over 2}}^{{\cal N}=1}\oplus  \left\{  S-{3\over 2} \right\}_{t+{3\over 2}}^{{\cal N}=1}  \oplus\left\{  S-1 \right\}_{t+2}^{{\cal N}=1} \,.  \ee

\textbf{Spin $S\geq 3$, $t=0$ PM multiplet:} 
\be \left\{S-1\right\}_{1,0}^{{\cal N}=2}= \left\{S-1\right\}_{1}^{{\cal N}=1}\oplus \left\{S-{1\over 2}\right\}_{{3\over 2}}^{{\cal N}=1}\oplus\left\{S-{3\over 2}\right\}_{{3\over 2}}^{{\cal N}=1}\oplus\left\{{S-1 }\right\}_{{ 2}}^{{\cal N}=1}. \ee
The extended module and alternate quantized spin $S-1$ go into $ \left\{S-1\right\}_{1}^{{\cal N}=1}$.  Apart from this we have only PM multiplets on the ${\cal N}=1$ side.

\textbf{Spin $S\geq 7/2$, $t=1/2$ PM multiplet:}  Splits into four different partially massless ${\cal N}=1$ multiplets,
\be  \left\{  S-1 \right\}_{{3\over 2},0}^{{\cal N}=2}= \left\{  S-1 \right\}_{{3\over 2}}^{{\cal N}=1} \oplus  \left\{  S-{1\over 2} \right\}_{2}^{{\cal N}=1}\oplus  \left\{  S-{3\over 2} \right\}_{2}^{{\cal N}=1}  \oplus\left\{  S-1 \right\}_{{5\over 2}}^{{\cal N}=1} \,.  \ee

\bigskip

\section{ ${\cal N}>2$ Supersymmetry}

We now present a heuristic argument that higher ${\cal N}$ supersymmetries should admit short, non-unitary multiplets containing partially massless fields of depth $t=S-{\cal N}$ where, as before, $S$ is the spin of the partially massless field (in contrast to the spin of the superconformal primary).  We follow the same general argument as is presented in Appendix \ref{normsappendex}.

Let us consider the quadratic Casimir operator of the superconformal algebra as it acts on a superconformal primary with zero $r$-charge.  Extending the ${\cal N} = 2$ case \cite{Bobev:2015jxa}, we find
\be
{\cal C}_2^{super} =D^2 + J_iJ_i- \frac{1}{2}\{P_i,K_i\} + \frac{1}{4} [S^{a I},Q_a^{~I}] + \ldots \, ,
\ee
where $I = 1,\ldots,{\cal N}$ and the $\ldots$ in the above expression denote operators that vanish on a primary state of zero $r$-charge.  Acting on an uncharged superconformal primary, we thus have
\be
{\cal C}_2^{super} \ket{\Delta, 0}^{a_1...a_{2s}} = \left[ \Delta(\Delta-3+{\cal N})+s(s+1) \right]\ket{\Delta, 0}^{a_1...a_{2s}} \, .
\ee
We expect multiplet shortenings to occur when a descendent state is itself a superconformal primary, i.e., when $S^{a I}\ket{\Delta', r'}^{a_1...a_{2s'}} = 0$ for the descendent state.  Let us consider the lowest spin state in the supermultiplet: for a superconformal primary given by $\ket{\Delta, 0}^{a_1...a_{2s}}$, this state will have quantum numbers $\Delta' = \Delta+\tfrac{{\cal N}}{2}$, $s'=s-\tfrac{{\cal N}}{2}$ and $r=0$ (see, e.g., \cite{Cordova:2016emh}).  If this state is also a superconformal primary then its quantum numbers must obey
\be
\label{SUSYcas}
\Delta(\Delta-3+{\cal N})+s(s+1)= \Delta'(\Delta'-3+{\cal N})+s'(s'+1) \, .
\ee
Solving gives $\Delta=s-{\cal N}+2$.  We now consider the highest spin conformal primary in the supermultiplet with quantum numbers $\Delta'' = \Delta+\tfrac{{\cal N}}{2}$, $S \equiv s''=s+\tfrac{{\cal N}}{2}$ and $r=0$.  This gives the condition $\Delta'' = S-{\cal N}+2= t+2$.  We see that the highest spin state in the candidate short supermultiplet is partially massless particle of depth $t=S-{\cal N}$.  For ${\cal N}=1$ this is simply the usual unitary massless representation.  For ${\cal N}=2$ we have the depth $t=s-2$ representations found above which include the partially massless spin-2 particle.

Allowing for non-zero $r$-charge and based on the ${\cal N} = 1$ and ${\cal N}=2$ cases, we might extrapolate to the case of general ${\cal N}$.   We would predict that in fact, higher ${\cal N}$ supersymmetries should admit short, non-unitary multiplets containing partially massless fields of depths $t=S-{\cal N}, \dots, S-1$.  We can anticipate that the generic ${\cal N}$-extended partially massless representations will be ${\cal N}\times {\cal N}$ diamonds in Figure \ref{allPM2fig}, with the ${\cal N}$ types of shortening (including the massless case) happening as the diamond approaches the massless line from below.

\bigskip
\section{Conclusions}

In this paper we have extended our earlier work \cite{Garcia-Saenz:2018wnw} on supersymmetric versions of partially massless fields by analyzing the case of extended SUSY with $\mathcal{N}=2$ supercharges. Our results go beyond partially massless particles: they provide new non-unitary the representations of the 3-dimensional $\mathcal{N}=2$ superconformal algebra, and hence of the equivalent super-AdS$_4$ algebra, of which partially massless SUSY multiplets are a special case. We have found a very rich set of possibilities for the structure of non-unitary representations that have no analogues in the unitary domain. These include a range of short multiplets and extended multiplets that differ qualitatively from the known unitary ones.  All shortening and extended multiplet conditions are summarized in Figs.\ \ref{shortsplane0}, \ref{shortsplane1half} and \ref{shortsplanes}.

Concerning the $\mathcal{N}=2$ representations that include PM fields, we have shown that the corresponding multiplets can be either long or short, unlike what occurs for $\mathcal{N}=1$ SUSY where PM particles always live in multiplets with no null states (with the exception of the exactly massless case). The generic long PM multiplet is given in \eqref{PMmultipletpicse5}. This multiplet is interesting in that it includes only gauge fields and yet no shortening occurs, something that doesn't exist in the unitary region. For lower spins and for special values of the PM depth we find several short multiplets, which can be fully PM or also include massive particles. Of special interest is the multiplet shown in \eqref{PMmultipletpicse7} which contains a single PM spin-2 field and no higher-spin states, and also involves an extended module of spin-1 states.

There are several generalizations of our work that might be worth investigating.   One is a detailed extension to dS space.
As mentioned in the introduction, the $\mathcal{N}=2$ AdS multiplets should have counterparts in dS spacetime. However the properties of the dS versions of our non-unitary representations, notably the ones with PM states, are unknown at the moment. It would be interesting to work out explicit dS Lagrangian formulations for these multiplets, and in particular to see what the AdS extended modules transmute to and if some analogue of this phenomenon exists in dS.  A more complete understanding of supersymmetric Lagrangians for PM particles, both in AdS and dS, would also open the door to the study of interactions. This was in fact one of our original motivations to study SUSY multiplets containing PM fields, given the obstacles encountered in the case of field multiplets with only spin-2 particles \cite{Deser:2013uy,deRham:2013wv,Garcia-Saenz:2014cwa,Joung:2014aba,Cherney:2015jxp,Garcia-Saenz:2015mqi}. That being said, the exploration of interactions via non-Lagrangian methods is also an intriguing possibility.

It would also be interesting to find Lagrangian descriptions for the supermultiplets.  Lagrangians for some lower spin ${\cal N}=1$ cases can be found in \cite{Garcia-Saenz:2018wnw}, and more complete results for Lagrangians of all spins in a frame-like gauge invariant description can be found in \cite{Buchbinder:2019olk}.  We have not attempted to construct Lagrangians realizing ${\cal N}=2$ SUSY on PM fields, but this would be interesting because of the new phenomena of extended multiplets that occur, whose Lagrangian description is unclear.
(Other previous work on Lagrangian formulations of partially massless and supersymmetric high spin fields includes \cite{Ondo:2016cdv,Zinoviev:2018juc,Buchbinder:2019dof,Buchbinder:2019olk,Khabarov:2019dvi,Khabarov:2020glf,Buchbinder:2020rex,Bachas:2019rfq}.)

Although a detailed analysis of SUSY multiplets with $\mathcal{N}>2$ would be messy, it would be worthwhile to have a better grasp on their generic properties, and we have taken only a first step here by showing some basic features of PM multiplets with higher $\mathcal{N}$. Perhaps more intriguing would be to explore supersymmetric extensions of PM fields beyond $D=4$ dimensions. It is known that some cubic vertices for PM spin-2 particles can be constructed when $D=4$ \cite{Zinoviev:2006im,Hassan:2012gz,Hassan:2012rq,deRham:2012kf,Hassan:2013pca,Hassan:2015tba,Apolo:2016ort,Boulanger:2018dau,Joung:2019wwf,Boulanger:2019zic}, which makes this value of the dimension somewhat special, and it would therefore be interesting to see if analogous constraints exist in the supersymmetric context.

Finally, it would be interesting to see how the electromagnetic duality of partially massless fields, which is present for both bosons \cite{Deser:2013xb,Hinterbichler:2014xga,Hinterbichler:2016fgl,Boulanger:2018shp,Boulanger:2018adg} and fermions \cite{Deser:2014ssa}, interacts with SUSY.

\bigskip
{\bf Acknowledgements:}
The authors are grateful to Leo Stein for helpful correspondence and Mathematica code. SGS is supported by the European Union's Horizon 2020 Research Council grant 724659 MassiveCosmo ERC-2016-COG; he would also like to thank Columbia University for hospitality when this work was initiated. KH acknowledges support from DOE grant DE-SC0019143 and Simons Foundation Award Number 658908. RAR is supported by DOE grant DE-SC0011941.  Both NB and RAR are supported by Simons Foundation Award Number 555117.

\bigskip 

\appendix

\section{Finding norms for $s>1$\label{normsappendex}}

Here we present some details on the computations of the norms of the conformal primaries within the superconformal multiplets $\{s\}_{\Delta,r}$.
For $s=0$ in Section \ref{s0normsubsec}, $s=1/2$ in Section \ref{s1/2normsubsec} and $s = 1$ in Section \ref{s1normsubsec}, we have computed the norm of each conformal primary explicitly.  
However, this brute force calculation of the norms becomes prohibitively involved at higher spin.  Instead, we make reasonable assumptions to extrapolate the norms of the multiplets at arbitrary $s$, up to an overall constant of determined sign.  

First, we exploit the fact that each of the conformal primaries in the superconformal multiplet will have the same eigenvalue of the superconformal quadratic Casimir operator.  In $d=3$, the quadratic conformal Casimir is given by 
\be
{\cal C}_2 =  D^2 + J_iJ_i- \frac{1}{2}\{P_i,K_i\}=D(D-3)+ J_iJ_i-P_iK_i \, .
\ee
For the ${\cal N}=2$ superconformal algebra, the quadratic Casimir gets extended to \cite{Bobev:2015jxa}
\be
{\cal C}_2^{super} = {\cal C}_2 + \frac{1}{4} [S^{a I},Q_a^{~I}] -\frac{1}{2}R^2+ \ldots \, .
\ee
The $\ldots$ denotes operators that vanish on a superconformal primary $\ket{\Delta, r}^{a_1...a_{2s}}$, i.e., a spin-$s$ conformal primary with conformal weight $\Delta$, which is also an eigenstate of the $R$-symmetry generator as in \eqref{scprsymmee}
and which is also annihilated by $S^{a I}$.
The quadratic Casimir acting on a superconformal primary becomes
\be
{\cal C}_2^{super}\ket{\Delta, r}^{a_1...a_{2s}} = \left[ \Delta(\Delta-1)+s(s+1)-\tfrac{1}{2}r^2  \right]  \ket{\Delta, r}^{a_1...a_{2s}} \, .
\ee

We expect shortenings to occur when a descendent state is itself a superconformal primary, i.e., when $S^{\alpha I}\ket{\Delta', r'}^{a_1...a_{2s'}} = 0$ for the descendent state, and thus this state decouples from the original multiplet.  In order for this to be true, given the original superconformal primary $\ket{\Delta, r}^{a_1...a_{2s}}$, there must exist a descendent state $\ket{\Delta', r'}^{a_1...a_{2s'}}$ with eigenvalues such that
\be
\label{SUSYcas}
\Delta(\Delta-1)+s(s+1)-\tfrac{1}{2}r^2 = \Delta'(\Delta'-1)+s'(s'+1) -\tfrac{1}{2}r'^2\, .
\ee
We can solve this equation for each conformal primary in the superconformal multiplet to find all possible new shortening conditions at each level.

We note that the superconformal multiplets do not necessarily take advantage of each of these shortening conditions.  For the $s=0$ multiplet there is no $\Delta = 0$ shortening condition at level 2 and no new shortening at level 3; for the $s=\tfrac{1}{2}$ multiplet there is no $\Delta = 1\pm r$ shortening condition at level 2 and there are no new shortening conditions at either level 3 or level 4; for the $s=1$ multiplet there is also no $\Delta = 1\pm r$ shortening condition at level 2 and no new shortening conditions level 3.  We can predict when a conformal primary exhibits new shortening by looking at the structure of the primary itself. First, we remark that if a level-$N$ conformal primary $\ket{P}_{N}$ is the $Q$-descendent of $\ket{P}_{N - 1}$ which shortens at $\Delta_{\star}$, then $\ket{P}_{N}$ must also shorten at $\Delta_{\star}$, unless $\Delta_{\star}$ appears in the denominator of a term in $\ket{P}_{N}$, as in the case of $\Delta_{\star} = 0$ at level 4 in the generic superconformal multiplet.  Moreover, there is maximal number of shortenings which a conformal primary may admit. For example, the norm of the $[s + 1]_{\Delta + 1, r}$ conformal primary
\begin{equation}
    \bar{Q}^{(a}Q^{b}\ket{\Delta, r}^{a_1...a_{2s})}  - \Big(\frac{\Delta - r + s}{\Delta + s}\Big)P^{(a b}\ket{\Delta, r}^{a_1...a_{2s})}
\end{equation}
is a third degree polynomial in $\Delta$, and thus admits 3 shortenings.  It is the $Q$-descendent of either $[s + 1]_{\Delta + \frac{1}{2}, r\pm1}$, which uses two of the available shortenings.  This leaves room for one new shortening, which is predicted by the Casimir.  Generically however, the norm of a generic conformal primary may not have enough roots to host the shortenings of the previous levels, as well as a new shortening condition. In such cases, the predicted shortenings of the Casimir are not realized.  In this manner, we can predict the numerator of the norm of conformal primary, up to an overall factor.  We can predict the denominator of the norm by multiplying together all of the denominators which appear in the coefficients of the $P$-descendants in the conformal primary.  This leaves only an overall undetermined constant.  The sign of this constant can be fixed by requiring that the norms be positive for large $\Delta$.  

We emphasize that we are merely conjecturing the structure of the norm for conformal primaries with $s\geq1$.  There may be for instance, accidental cancellations, so that not all of the denominators in a conformal primary make an appearance in the norm.  For example, this happens at level 4 in the $\{\frac{1}{2}\}_{\Delta, r}$ multiplet.  However, we have explicitly computed all of the norms for $s = 1$ and our conjecture correctly predicts all of them.  For generic $\Delta$, the $s = 1$ case should be generic, so we take this as good evidence that our conjectured norms are valid.

\bigskip

\bibliographystyle{utphys}
\addcontentsline{toc}{section}{References}
\bibliography{N2_arxiv}

\end{document}